# Honing Theory: A Complex Systems Framework for Creativity

**Liane Gabora,** *University of British Columbia*

RUNNING HEAD: [Complex Systems Framework Creativity]

Correspondence should be addressed to:

Liane Gabora

Department of Psychology, University of British Columbia

Okanagan Campus, 1147 Research Road, Kelowna BC, Canada V1V 1V7

Email: liane.gabora@ubc.ca

Tel: 604-358-7821 / 250-807-9849

*Abstract: This paper proposes a theory of creativity, referred to as honing theory, which posits that creativity fuels the process by which culture evolves through communal exchange amongst minds that are self-organizing, self-maintaining, and self-reproducing. According to honing theory, minds, like other self-organizing systems, modify their contents and adapt to their environments to minimize entropy. Creativity begins with detection of high psychological entropy material, which provokes uncertainty and is arousal-inducing. The creative process involves recursively considering this material from new contexts until it is sufficiently restructured that arousal dissipates. Restructuring involves neural synchrony and dynamic binding, and may be facilitated by temporarily shifting to a more associative mode of thought. A creative work may similarly induce restructuring in others, and thereby contribute to the cultural evolution of more nuanced worldviews. Since lines of cultural descent connecting creative outputs may exhibit little continuity, it is proposed that cultural evolution occurs at the level of self-organizing minds; outputs reflect their evolutionary state. Honing theory addresses challenges not addressed by other theories of creativity, such as the factors that guide restructuring, and in what sense creative works evolve. Evidence comes from empirical studies, an agent-based computational model of cultural evolution, and a model of concept combination.*

**An Integrative Honing Framework for Creativity**

Creativity is central to cognition, and one of our most human traits. It plays an important role in abilities such as planning, problem solving, and story telling, and has given rise to art, science, and technology. It allows us to imagine beyond the present to reconstruct the past or fantasize about the future. The Mona Lisa, the roller coaster, and the stock market all reflect the ingenuity of the human mind. Our capacity to innovate, build on one another's inventions, and adapt them to our own needs and tastes, has transformed this planet.

Creativity refers to the process by which new and valued or appropriate outputs are generated (e.g., inventions or poems), and individuals who generate such outputs are said to be creative. Over the last half-century studies have revealed that creativity is correlated with personality traits such as norm-doubting, tolerance of ambiguity, and openness to experience (Barron, 1969; Batey & Furnham, 2006; Eysenck, 1993; Feist, 1998; Martindale & Daily, 1996), and with activation of particular brain networks (Vartanian, Bristol, & Kaufman, 2013) and related in interesting ways to culture (Lubart, 1990), family birth order (Bliss, 1970; Sulloway, 1996), and a sense of complete absorption or 'flow' (Csikszentmihalyi, 1996). However, although we have gained much in the way of fragmentary knowledge about creativity, an integrated framework for creativity eludes us (Bowden, Jung-Beeman, Fleck, & Kounios, 2005; Sternberg & Kaufman, 2010). Though formal models of creative cognition have been around for some time (e.g., Langley, Simon, Bradshaw, & Zytkow, 1987), we do not know how to formally describe a 'half-baked' idea, or why immersion in creative tasks can be transformative or even therapeutic, or why humans are so much more creative than other species.

The goal of this paper is to propose a psychologically, neurologically, and evolutionarily plausible complex, adaptive systems (CAS) framework for how creative ideas unfold over time, both in the minds of individuals, and through interaction amongst individuals. The theory of creativity proposed here grew out of the attempt to develop a rigorous theory of how culture evolves (Gabora, 1995, 2004, 2008a, 2013), building on earlier ideas about the relationship between complexity theory and creativity (e.g., Guastello, 2002; Orsucci, 2002; Richards, 1996; Schuldberg, 1999), and the role of creativity in fueling historical change (e.g., Dasgupta, 1994; Martindale, 1994). Humans possess two levels of complex, adaptive, self-organizing, evolving structure; in addition to an organismic level they have a psychological level (Barton, 1994; Combs, 1996; Freeman, 1991; Pribram, 1994; Varela, Thompson, & Rosch, 1991). It is proposed that this psychological level enables them to detect gaps or inconsistencies and consider them from different perspectives such that they may find a form that fits with their models of the world, such that these mental models become more robust. Creative thinking induces *restructuring* of representations, which may involve re-encoding the problem such that new elements are perceived to be relevant, or relaxing goal constraints (Weisberg, 1995). However, according to the theory of creativity introduced here, referred to as honing theory (HT), the transformative impact of immersion in the creative process extends far beyond the "problem domain"; it can bring about sweeping changes to that second (psychological) level of complex, adaptive structure that alter one's self-concept and view of the world. Creative acts or products can then render such cognitive transformation culturally transmissible. Thus minds evolve through culture, and creative outputs are external manifestations of this process. The paper examines evidence for HT from research on associative memory, empirical studies, and computational experiments. The bulk of the paper addresses the process by which an intrinsically motivated individual creates, i.e., the cognitive process culminating in the execution of a creative idea. It does not provide an integrated account in the sense of incorporating all of what are sometimes referred to as the four P's of creativity (process, product, person, and place); the focus is almost exclusively on the creative process. Toward the end the paper addresses how a creative idea accumulates change as it is considered by one individual after another and spreads through a society.

## Existing Evolutionary Frameworks

Ideas clearly adapt over time, and since adaptation is the crux of evolution (Laszlo, 1991) it seems reasonable that there is a sense in which ideas *evolve*. Since this paper proposes an explanatory framework for how ideas evolve, it is necessary to briefly mention two other evolutionary approaches.

According to the Blind Variation Selective Retention (BVSR) theory of creativity, novelty comes about through trial and error; large numbers of distinct and separate ideational variants are generated sequentially through an essentially "blind" process, followed by selective retention of the fittest for development into a finished product (Campbell, 1960; Simonton, 1999). Evidence that the generation of creative variation is not blind but guided by factors such as remote associates led Simonton (2007, 2011) to a "partially sighted" version of BVSR (see Dietrich and Haider (2015) for a similar theory). A related notion also appears in the *structure mapping* theory of analogy (Gentner, 1983), according to which analogy generation occurs in two steps: first, searching memory in a "structurally blind" manner (Gentner, 2010, p. 753) for an appropriate source and aligning it with the target, and second, mapping the correct one-to-one correspondences between source and target.

BVSR has been referred to as a Darwinian theory of creativity (Simonton, 1999, 2007), as is Dietrich and Haider's (2015) evolutionary-predictive theory. The logic underlying a Darwinian theory of creativity is that ideas adapt over time, and Darwinian selection is the best-known means to accomplish this. A detailed discussion of the serious problems with the BVSR approach can be found elsewhere (e.g., Gabora, 2005, 2011a; Gabora & Kauffman, in press). In brief, to say that creativity is Darwinian is not to say that it consists of variation plus selection—in the everyday sense of the term; it is to say that evolution is occurring because selection is affecting the distribution of randomly generated heritable variation across generations. However, there is compelling evidence that variation is not random but chaotic (Guastello, 1995); moreover, in creative thought the distribution of variants is not key, i.e., one is not inclined toward idea A because 70% of one's candidate ideas are variants of A while only 30% are variants of B; one is inclined toward whichever seems best. Simonton has now "radically reformulated" BVSR such that there is no mention of randomness and its "explanatory value does not depend on any specious association with Darwin's theory of evolution" (Simonton, 2012, p. 48). Thus, this new incarnation does not appear to be an evolutionary theory of creativity.

Another possibility is that creative ideas evolve not in individual minds but amongst culturally interacting individuals. Indeed, human cultural change is adaptive (tends to result in a better fit between individuals and their environments), cumulative (elements of culture build on one another), and open-ended (the set of possible outputs cannot be predefined). Since natural selection is the most well known evolutionary process, it has been tempting to assume that cultural evolution is Darwinian, i.e., occurs by means of natural selection or a process that is analogous or algorithmically equivalent to it (Boyd & Richerson, 1985; Gabora, 1996; Mesoudi, Whiten, & Laland, 2006). However, as discussed in detail elsewhere (Gabora, 2004, 2011b, 2013; Tëmkin, & Eldredge, 2007), this theory suffers from many of the same problems as the Darwinian theory of creativity. Because organisms replicate by way of a self-assembly code there is no transmission of traits acquired over a lifetime (e.g., one does not inherit a parents' tattoo). Because elements of culture (e.g., tools or tunes) lack the algorithmic structure of a self-assembly code, cultural diffusion of acquired traits is widespread (e.g., if a cup acquires the trait of having a handle, subsequent cups may also have handles). Since acquired change can operate instantaneously whereas inherited change requires generations, non-negligible acquired change, its presence overwhelms any change due to selection acting on distributions of randomly generated heritable variation. Thus change over time reflects not selection on relative frequencies of variants but whatever is biasing the generation of variants in the first place (Fig. 1).

--------------------------------

Insert Figure 1 About Here

--------------------------------

Until a few decades ago it was believed that natural selection was the only way to evolve. The wealth of knowledge that has accumulated in recent decades concerning primitive, non-Darwinian means of adaptive evolution (as outlined below) suggest a new integrative framework for understanding creativity as the process that enables minds to restructure and self-organize, and thereby culture to evolve.

## HONING THEORY OF CREATIVITY

This paper will argue that the architecture of associative memory is conducive to restructuring and insight through tractable means that need not involve blind (or partially blind) sequential search amongst discrete variants. Like the Geneplore model (Finke, Ward, & Smith, 1992), HT is consistent with the notion that creativity involves the generation and exploration of 'pre-inventive structures'. However, HT attempts to formalize the notion of a pre-inventive structure, and is not committed to the view that there need be more than one of them. Like Darwinian approaches (Dietrich & Haider, 2015; Simonton, 1999), it seeks to explain in what sense ideas evolve over time. However, HT is not a Darwinian theory of creativity; it does not posit that creativity arises because of the effect of selection over generations on the distribution of heritable variation across a population.

Creativity is often thought to originate with the generation or awareness of what Torrance (1962) calls a *gap*, or sense of incompletion, which may arise spontaneously, or slowly over the course of years, and be trivial (e.g., drawing a squiggle and wondering what additional doodling would make the squiggle appealing) or of worldly consequence. The gap that precipitates creative thinking may be described as a (relatively) chaotic cognitive state (Guastello, 2002), and be accompanied by a lingering feeling that compels the exploration and expression of ideas (Feinstein, 2006). The notion of a gap implies the existence of a larger conceptual structure generally assumed to be the "problem domain". However, ingredients for creative thinking, such as remote associates (Mednick, 1962), analogies (Gentner, 1983), and inspirational sources or 'seed incidents' (e.g., being in love inspires a song) (Doyle, 1998; Gabora & Carbert, 2015), often come from *beyond* the problem domain. This observation serves as a departure point for HT, which asserts that since the contents of the mind collectively self-organize into a loosely integrated, structure, none are a priori excluded from the creative task. The notion that the mind can self-organize into a new patterned structure of relationships has been around for some time (Ashby, 1947; Barton, 1994; Carver & Scheier, 2002; Guastello, 2002), as has the idea that this facilitates creativity (e.g., Abraham, 1996; Goertzel, 1997; Guastello, 1998). What has not been fully appreciated is that (a) creativity can exert widespread transformative effects on the mind, and therefore (b) our understanding of both creativity and cultural evolution will remain fragmented until we develop a framework for creative cultural change that gives central place to *how the mind self-organizes* in response to (1) the creative act, and (2) communal exchange with others and exposure to their creative outputs. The necessity of such a framework is particularly glaring when one considers phenomena that clearly entail cognitive restructuring beyond the immediate "problem domain", such as the transport of recognizable creative style from one domain to another (Gabora, O'Connor, & Ranjan, 2012; Ranjan, 2014), or findings that creators often cite sources outside their creative domain as the inspiration for their works (Carbert & Gabora, 2015). The interaction of information across domains may at least partly explain why people who specialize in one domain often make contributions in other domains as well (Guastello & Shissler, 1994).

**Creativity Fuels Cultural Evolution through Cognitive Restructuring**

The framework for creativity proposed here is based on arguments that the human mind is not just a complex, adaptive system but a unit of cultural evolution (Gabora, 2004, 2008a, 2013). At first glance it may appear that the basic units of cultural evolution are such things as rituals or tools. However, the self-organizing nature of the mind makes it impossible to trace all the influences or "conceptual parents" of a creative cultural contribution such as a piece of music or journal article; moreover, it is not clear how evolution is occurring at this level. For example, consider the situation in which a steam engine inspires a song, which inspires a book. The lack of continuity across this "line of descent" is too great for this to be productively construed as an evolutionary lineage. However, since the creators of these works live in the same world they share a considerable body of knowledge, and thus there *is* continuity at the level of the *worldview:* the web of beliefs, attitudes, ideas, procedural and declarative knowledge, and habits that emerge through the interaction between personality and experience. Thus, it is suggested that what evolves through culture is not isolated creative contributions but worldviews, and cultural contributions reflect the states of the worldviews that generate them.

Another argument for the position that it is worldviews that evolve through culture is that individual cultural contributions do not possess the algorithmic structure that Holland (1975) identified as necessary for evolution through a variation-selection process (Gabora, 2011a). Nor does a mind as a whole possess this structure. A mind, however, possesses the structure necessary to evolve through a more primitive non-Darwinian form of evolution involving self-organization of elements into a whole, sometimes referred to as *communal exchange* (Vetsigian, Woese, & Goldenfeld, 2006). Though the term 'communal exchange' appears to emphasize interaction *between* entities, it refers to a process in which interactions *within* entities play an equally important, and complementary role. Each entity in a communal exchange process is *self-organizing:* a stable global organization emerges through interactions amongst its parts (Prigogine &

Nicolis, 1977). It is also *autopoietic:* it not only maintains but also reproduces its own structure (Maturana & Varela, 1973). There is extensive evidence that for the first several million years of its existence life on earth evolved through communal exchange of components amongst self-organizing autocatalytic protocells, *prior to establishment of a genetic code*, and only once a RNA-based self-assembly code was in place did life evolve through natural selection (Gabora, 2006; Kauffman, 1993; Segre, 2000; Vetsigian et al., 2006; Wächtershäuser, 1997). Indeed, as outlined in 14 articles in Frontiers (Raoult & Koonin, 2012) and elsewhere (Lynch, 2007), the last century's emphasis on natural selection in evolutionary theory is not warranted by current research; even today, variation-selection is, quantitatively at least, not the dominant evolutionary process (Koonin, 2009). Like structures that evolve through variation-selection, structures that evolve through communal exchange are *self-regenerating;* they produce variants of themselves through duplication of their components (Farmer, Kauffman, & Packard, 1986; Hordijk, 2013; Kaufman, 1993). The fidelity of this primitive form of replication is lower than that of replication with a self-assembly code (such as RNA or DNA) through variation-selection, but sufficient to result in cumulative, adaptive, open-ended evolution. Note that without a self-assembly code safeguarding fidelity of replication, a particular component may play a different role in different autocatalytic networks; it is the autocatalytic network as a collective whole that is evolving.

The concept of evolution through communal exchange amongst self-organizing, autopoietic structures is not new; it is a well-established theory of how the earliest stage of biological life evolved. What is new is the proposal that the cultural evolution of human worldviews occurs through this kind of communal exchange process and that creativity is what it is fuels it. If even biological life initially evolved through non-Darwinian communal exchange it makes sense that cultural change would also not jump straight to the relatively complex Darwinian form of evolution, but (at least initially) evolve through the simpler but less efficient process of communal exchange amongst self-organized structures, in this case minds. Central to HT is the idea that minds evolve through communal exchange and self-organization mediated by what Hirsh, Mar, and Peterson (2012) refer to as *psychological entropy*. The concept of entropy, which comes from thermodynamics and information theory, refers to the amount of uncertainty and disorder in a system. Self-organizing systems continually interact with and adapt to their environments to minimize internal entropy. Open systems such as living organisms capture energy (or information) from their environment, use it to maintain semi-stable, far-from-equilibrium states, and displace entropy into the outside world to keep their own entropy low. (The displaced entropy is sometimes called negentropy.) Hirsh et al. use the term psychological entropy to refer to anxiety-provoking uncertainty, which they claim humans attempt to keep at a manageable level. Here we expand on this slightly; we use the term to refer to *arousal*-provoking uncertainty, which can be experienced negatively as anxiety or positively as a wellspring for creativity (or both). Redefining psychological entropy in terms of arousal rather than anxiety is consistent with findings that creative individuals exhibit greater openness to experience and higher tolerance of ambiguity (Feist, 1998), which could predispose them to states of uncertainty. Their higher variability in arousal reflects a predisposition to invite situations that increase psychological entropy, experience them positively, and resolve them. It is proposed that creativity uses psychological entropy—a macro-level variable acting at the level of the worldview as a whole—to drive emotions and intuitions that play a role in tracking and monitoring creative progress. This is supported by findings that insight—defined as a sudden new representation of a task or realization of how to go about it (Mayer, 1995), or conscious cognitive reorganization culminating in a non-obvious idea or interpretation (Kounios & Beeman, 2009, 2014)—tends to be accompanied by feelings of surprise (Boden, 1990) and confidence (sometimes mistakenly) in the truth or correctness of the idea (Gick & Lockhart, 1995). Using psychological entropy to guide what Hummel and Holyoak (1997) refer to as self-supervised learning, the creator hones ideas by viewing them from new perspectives.

Psychological entropy—experienced as dissonance, boredom, curiosity, or a problem—may signal that a particular arena of understanding could benefit from self-organized restructuring. Just as a wounded organism spontaneously heals, when one encounters a situation that seems unjust, or that challenges expectations or beliefs, there is cognitive dissonance, and a spontaneous attempt to weave a story, fortify knowledge, or revise beliefs to accommodate the challenge (Antrobus, Singer, Goldstein, & Fortgang, 1970; Festinger, 1957; Klinger & Cox, 1987). We can use the term *self-mending* to refer to the capacity to reduce psychological entropy in response to a perturbation, i.e., it is a form of self-organization involving the reprocessing of arousal-provoking material. A worldview (like an organism) is self-mending to the extent that one is inclined to explore possibilities and revise interpretations to establish or restore consistency or integrity (Gabora & Merrifield, 2012; Greenwald, Banaji, Rudman, Farnham, Nosek, & Mellott, 2002; Osgood & Tannenbaum, 1955).[2]

A worldview is restructured through the strengthening and weakening of existing associations, and the formation of new ones. There is no need for memory to be searched or randomly sampled for creative associations to be made. The fact that representations are *distributed* across cell assemblies of neurons that are sensitive to particular high-level or low-level features ensures the feasibility of forging associations amongst items that are related (Rumelhart & McClelland, 1986), perhaps in a surprising but useful or appealing way. Neurons exhibit *coarse coding:* although each neuron responds maximally to a particular microfeature it responds to a lesser degree to similar microfeatures. Memory is also *content addressable;* there is a systematic relationship between stimulus content and the cell assemblies that encode it, such that memory items are evoked by stimuli that are similar or 'resonant' (Hebb, 1949). This enables reinterpretation of higher order relations between perceptual stimuli through synchronization of prefrontal neural populations (Penn, Holyoak, & Povinelli, 2008). The distributed, content-addressable nature of memory is critical for creativity (Gabora, 2000, 2010). The more detail with which items are encoded in memory, the greater their overlap, and the more routes for creatively re-interpreting experience. If there is overlap between regions where distributed representations are encoded they can 'find' one another. They may have been encoded at different times, under different circumstances, and the relationship between them never explicitly noticed, but some situation could make their relationship apparent.

A worldview is not just self-organizing and self-mending but *self-regenerating:* an adult shares experiences, ideas, and attitudes with children (and other adults), thereby influencing the conceptual closure process by which other worldviews form and transform. Children expose fragments of what was originally the adult's worldview to different experiences, different bodily constraints, and thereby forge unique internal models of the relationship between self and world. Because there is no self-assembly code safeguarding fidelity of replication, a particular belief or idea may play a different role in different worldviews. Thus, a worldview evolves through communal exchange by interleaving (1) internal interactions amongst its parts, and (2) external interactions with others. It not only self-organizes in response to perturbations but the whole is imperfectly reconstituted and passed down through culture. The more information it encodes, the more ways it can be configured. The state of a worldview reveals itself through behavioural regularities in how it creatively expresses itself and responds to situations.

Although selection as the term is used in everyday parlance may play a role (individuals may be selective about which aspects of their worldviews they share or assimilate, or, for example, which paintings they show at a gallery), the process does not involve selection in its technical sense (change over generations due to the effect of differential selection on the distribution of heritable variation across a population). HT posits that, rather than search, creativity involves viewing the creative task from a new context, which may restructure the internal conception of the task and modify the external product. This in turn suggests a new context, recursively, until the product reaches an acceptable form. HT is compatible with Sabelli and Abouzeid's (2003) claim that creativity is characterized by (1) *diversification,* indicating an expanded state space (in contrast to processes that converge to equilibrium or to periodic or chaotic attractors), (2) *novelty,* i.e., less recurrence than obtains from random series, and (3) *rearrangement*, a measure of patterned recurrences that indicates nonrandom complexity.

**Development of the Capacity for Creativity**

To contribute to culture in a way that differs significantly from what has come before, it is helpful to be able to think about existing ideas in a new light by engaging in abstract thought. However, abstract thought is the process that connects ideas in the first place, i.e., puts them within reach of one another. Thus, to explain how a complex system composed of mutually interdependent parts comes into existence we have a chicken and egg problem.

A tentative solution (Gabora, 1998) was inspired by Kauffman's (1993) application of developments in graph theory (Erdos & Renyi, 1960) to the question of how life began. Kauffman recognized that origin of life theories must solve the "chicken and egg" problem of how a complex system composed of mutually dependent parts comes into existence. He showed that when catalytic polymers interact their diversity increases, as does the probability that some subset of the total reaches a percolation threshold at which point there is a catalytic pathway to every member, a state referred to as autocatalytic closure. He demonstrated that autocatalytic sets emerge for a wide range of hypothetical chemistries—i.e., different collections of catalytic polymer molecules. Each of the original "food set" polymers was composed of up to a maximum of *M* monomers, and assigned a low a priori probability *P* of catalyzing the reaction by which another polymer was formed. Some sets were subcritical—unable to incorporate new polymers. Others were supra-critical—they incorporated new polymers so often that their behavior was chaotic. Still others incorporated new polymers with sufficient frequency to obtain closure and exist at the proverbial edge of chaos. Which of these regimes a set fell into depended on *M* and *P*. The lower the value of *P*, the greater *M* had to be, and *vice versa,* for closure to occur.

Let us examine how this applies to the process by which a child becomes a potentially creative contributor to culture. The analog of catalytic molecules is items encoded in memory. The analog of *M*, the maximum polymer length, is the maximum number of features of an item in memory. The analog of *P*, the probability of catalysis, is the probability that one representation becomes associated with or dynamically bound to another through synchronized activity amongst prefrontal neural populations. Exposure to highly similar items prompts the formation of abstract concepts that encompass these instances. Eventually the memory reaches a percolation threshold at which point the number of ways of forging associations amongst memory items increases exponentially faster than the number of them. Interactions amongst memory items increases their joint complexity, transforming them into a network, such that any new stimulus can be reinterpreted in terms of previous knowledge and experience. This enables the creative connecting and refining of ideas necessary for the individual to become a participant in the evolution of cultural novelty. Since this is an entropy-minimizing process the worldview achieves a more stable state that (following Kauffman's use of the term 'autocatalytic closure') has been referred to as *conceptual closure* (Gabora, 2000). If the probability of associative recall is low the network is subcritical; it has too much difficulty assimilating new information to become an integrated conceptual web. If the probability of associative recall is high the network is super-critical; it assimilates new information readily but risks destabilization. In conceptual closure, as was the case with autocatalytic closure, if either *M* or *P* is too high the network may be too chaotic to establish and maintain closure. Since systems—including minds—that are in or near chaotic regimes tend to exhibit creative behavior (Guastello, 2002; Richards, 1996; Schuldberg, 1999), creativity is expected to be highest when *M* and *P* fall just shy of super-criticality.

Thus a model exists for how an autopoietic structure comes into existence, and HT posits that applying this model to the mind is essential to understanding creativity. The network can thereafter change and expand to incorporate roles, fillers, and hierarchically structured relations, using neural synchrony and dynamic binding to grow and embellish the basic network structure. The network may capitalize on the opportunities of a fine-grained associative memory for creative combinations without straying permanently from the critical regime between subcritical and supercritical by shifting between different modes of thought. Psychological theories of creativity typically involve a divergent stage that predominates during idea generation and a convergent stage that predominates during the refinement, implementation, and testing of an idea (for a review see Runco, 2010; for comparison between divergent / convergent processes in creativity and dual process models of cognition see Sowden, Pringle, & Gabora, 2015). Divergent thought is characterized as intuitive and reflective, and as involving the generation of multiple discrete, often unconventional possibilities, possibly due to reduced latent inhibition (Carson, Peterson, & Higgins, 2003). It is contrasted with convergent thought, which is critical and evaluative, and as involving selection or tweaking of the most promising possibilities. There is empirical evidence for oscillations in convergent and divergent thinking, and a relationship between divergent thinking and chaos (Guastello, 1998). It is widely believed that divergent thought involves defocused attention and associative processing, and this is consistent with the literal meaning of divergent as "spreading out" (as in a divergence of a beam of light). However, the term divergent thinking has come to refer to the kind of thought that occurs during creative tasks that involve the generation of multiple solutions, which may or may not involve defocused attention and associative memory. Moreover, in divergent thought, the associative horizons simply widen generically instead of in a way that is tailored to the situation or context (Fig. 2). Therefore, we will use the term *associative thought* to refer to creative thinking that involves defocused attention and context-sensitive associative processes, and *analytic thought* to refer to creative thinking that involves focused attention and executive processes. The capacity to shift between these modes of thought has been referred to as *contextual focus* (CF) (Gabora, 2003). Thus while some dual processing theories (e.g., Evans, 2003) make the split between automatic and deliberate processes, CF makes the split between an associative mode conducive to detecting relationships of correlation and an analytic mode conducive to detecting relationships of causation. Defocusing attention facilitates associative thought by diffusely activating a broad region of memory, enabling obscure (though potentially relevant) aspects of a situation to come into play. Focusing attention facilitates analytic thought by constraining activation such that items are considered in a compact form that is amenable to complex mental operations.

-------------------------------

Insert Figure 2 About Here

-------------------------------

A plausible neural mechanism for CF has been proposed (Gabora, 2010). In a state of defocused attention more aspects of a situation are processed, the set of activated features is larger, and thus the set of possible associations is larger. Activation flows from specific instances, to the abstractions they instantiate, to other seemingly unrelated instances of those abstractions. Cell assemblies that would not be activated in analytic thought but that would be in associative thought are referred to as *neurds*. Neurds respond to features that are of marginal relevance to the current thought. They do not reside in any particular region of memory; the subset of cell assemblies that count as neurds shifts depending on the situation. For each different perspective one takes on an idea, a different group of neurds participates.

Neurds may generally be excluded from activated cell assemblies, becoming active only when there is a need to break out of a rut. In associative thought, diffuse activation causes more cell assemblies to be recruited, including neurds, enabling one thought to stray far from the preceding one while retaining a thread of continuity. Thus the associative network is not just penetrated deeply, but traversed quickly. There is greater potential for overlapping representations to be experienced as wholes resulting in the uniting of previously disparate ideas or concepts. While the preparation phase of the creative process likely involves long-term change to how ideas are encoded in the neocortex, the merging of thoughts culminating in insight—particularly when this involves generalization and inference triggered by a particular recent experience—may involve recurrent connections in the hippocampus (Kumaran & McClelland, 2012). Research on the neuroscience of insight suggests that alpha band activity in the right occipital cortex causes neural inhibition of sensory inputs, which enhances the relative influence of internally derived ("non-sensory") inputs, and thus the forging of new connections (Kounios & Beeman, 2014). Following insight, the shift to an analytic mode of thought could be accomplished through decruitment of neurds. Findings that the right hemisphere tends to engage in coarser semantic coding and have wider neuronal input fields than the left (Kounios & Beeman, 2014) suggest that the right hemisphere may predominate during associative thought while the left predominates during analytic thought. In short, the architecture of associative memory is conducive to using psychological entropy to guide self-supervised learning and achieve a more coherent worldview with creative output as an external manifestation of this process.

Creative individuals may be adept at adjusting *P* in response to the situation and thereby engaging in CF, tending toward associative thought when stumped and analytic thought for fine-tuning (Gabora, 2003). With the diffuse activation characteristic of associative thought, items are evoked in detail, or multiple items with overlapping distributions of features are evoked simultaneously. Indeed, diffuse activation is expected to result in the 'flat' associative hierarchies characteristic of highly creative people (Mednick, 1962). When activation is constrained, items are evoked in a compressed form, and few are evoked simultaneously. As such, constrained activation is expected to produce the 'spiky' associative hierarchies characteristic of uncreative people (Mednick, 1962). Spiky activation enhances ability to stay focused when accessing remote associations would be distracting, while diffuse activation is useful when conventional problem solving methods are ineffective. The capacity of creative individuals to shift between modes of thought may be related to greater variability in arousal, with higher resting states, and lower levels during creative activity (Martindale & Armstrong, 1974). This supports our use of the term psychological entropy to refer to arousal-provoking uncertainty.

This view of creativity is consistent with historiometric analyses of lifespan creativity. Although a few studies found no systematic relation between hit ratio (masterwork output divided by total output) and age (Simonton, 1977; Weisberg, 1994), others show a strong positive correlation, though creativity often peaks at approximately age 40 to 50 and decreases thereafter (e.g., Kozbelt, 2004, 2005, 2008; Zickar & Slaughter, 1999). Since with age an individual's worldview becomes richer and more complex but then tends to solidify, it becomes harder for the world to present something that provokes reconfiguration. Thus creative output is expected to diminish in latter life as is consistent with analyses of lifespan creativity.

**Characterizing Creative Cognitive States in terms of Potentiality**

The flip side of arousal-provoking uncertainty is *potentiality;* lack of knowledge about how an idea will unfold implies potential for different possible outcomes. In HT, an unfinished idea or pre-inventive structure is described not as a collection of possible thought trials but as a single state of potentiality. We now turn to a formal approach to modeling an idea in a state of potentiality that grew out of a theory of concept combination. Although many have proposed that concept combination lies at the heart of creativity (e.g., Finke, Ward, & Smith, 1992), few links have been made between theories of creativity and formal models of concept combination.

Since creative insight involves combining concepts and percepts and putting them in new contexts (Ward, Smith, & Vaid, 1997), a theory of creativity should arguably have as its foundation a theory of how they interact. Formal models of insight through concept combination have been proposed (Costello & Keane, 2000; Dantzig, Raffone, & Hommel, 2011; Thagard & Stewart, 2011). However, a difficulty that arises is that concepts are not just conjointly activated but interact in a highly contextual, almost chameleon-like manner that takes relational structure into account (Barsalou, 1999; Gagne & Spalding, 2009). Moreover, their interactions are non-compositional; conjunctions and disjunctions of concepts get used in ways that violate the rules of classical logic (Aerts, Aerts, & Gabora, 2009; Busemeyer & Bruza, 2012; Estes & Ward, 2002; Hampton, 1987; Kitto, Ram, Sitbon, & Bruza, 2011; Osherson & Smith, 1981; Rosch, 1973). For example, although people do not rate 'talks' as a characteristic property of PET or BIRD, they rate it as a characteristic property of PET BIRD. 'Talks' is not a property of either constituent concept; it is an emergent property of the combination. A combined concept may also exhibit loss of properties typical of one or both constituents. For example, 'surrounded by water', a seemingly central property of ISLAND, is (hopefully) not present in instances of KITCHEN ISLAND. Noncompositionality occurs with respect to not just properties but also exemplar typicalities, e.g., although people do not rate 'guppy' as a typical PET, nor as a typical FISH, they rate it as a highly typical PET FISH. Interestingly, people prefer concept combinations that interact non-compositionally over those that interact in an additive manner, and think they are more creative (Gibbert, Hampton, Estes, & Mazursky, 2012). This problem has stymied not just theories of concepts and creativity, but of cognition more generally, for the meaning and generativity of thought and language emerge through conceptual interaction.

The difficulty of modeling the behavior of concepts arises not just because they interact non-compositionally, but such interactions may give birth to completely new concepts, and incorporate world knowledge (Fodor, 1998; Hampton, 1987). Consider the concept TEAPOT, which presumably came about by taking a known object, a pot, and adapting it to be used for liquid. At the cultural level this invention built straightforwardly on previous knowledge. However, at the conceptual level, the combining of two concepts gave birth to a new concept, the idea of a spout. Moreover, this process employed knowledge of gravity and fluid dynamics that extends beyond the properties and exemplars of TEA and POTS.

The non-compositionality of conceptual interactions have made concepts resistant to mathematical description, a problem that must be solved to model how meaning emerges when people combine concepts and words into larger semantic units. Significant progress has been had modeling concepts using a generalization of the formalism of quantum mechanics (Aerts, 2009; Aerts & Gabora, 2005a,b; Aerts, Gabora, & Sozzo, 2013; Bruza, Kitto, Ramm, & Sitbon, 2011; Gabora & Aerts, 2002; Kitto et al., 2011). This approach is unrelated to the lines of research that use quantum models of the brain to understand consciousness (Hammeroff, 1998) or memory (Pribram, 1993) and makes no assumption that phenomena at the quantum level affect the brain; it draws solely on abstract formal structures that, as it happens, found their first application in quantum mechanics. The approach is, however, in some ways consistent with the proposal that quantum mechanics could provide a means of describing creative cognitive states that simultaneously incorporate multiple possibilities (Goswami, 1996; McCarthy, 1993).

Quantum probability models in psychology have been compared side-by-side with classical models (Busemeyer, Pothos, Franco, & Trueblood, 2011). According to classic probability, all events are subsets of a common sample space; that is, they are based on a common set of elementary events. An advantage of a quantum model over a classical model such as a Bayesian one (c.f. Arecchi, 2011) is that it uses variables and spaces that are defined specifically with respect to a particular context, which is necessary to model the contextual, compositional way they interact. The state $|\psi\rangle$ of an entity is written as a linear superposition of a set of basis states $\{|\phi_i\rangle\}$ of a Hilbert space $\mathcal{H}$, which is a complex vector space with an inner product.[3] Another advantage of a quantum model over a classical one is that it uses *amplitudes*, which though directly related to probabilities, can exhibit interference, superposition, and entanglement, which are also needed to capture certain aspects of how concepts behave (Aerts, 2009; Aerts, Aerts, & Gabora, 2009; Aerts, Broekaert, Gabora, & Veloz, 2012; Aerts, Gabora, & Sozzo, 2013; Aerts & Sozzo, 2011; Bruza, Kitto, Nelson, & McEvoy, 2008; Bruza et al., 2011; Nelson & McEvoy, 2007). The amplitude term, denoted *ai*, is a complex number that represents the contribution of a component state $|\phi_i\rangle$ to the state $|\psi\rangle$. Hence $|\psi\rangle = \Sigma_i a_i |\phi_i\rangle$. The square of the absolute value of the amplitude equals the probability that the state changes to that particular component basis state upon measurement. A non-unitary change of state is called *collapse*. The choice of basis states is determined by the observable $O$ to be measured, and its possible outcomes $o_i$. The basis states corresponding to an observable are referred to as *eigenstates*. Observables are represented by self-adjoint operators on the Hilbert space. The lowest energy state is referred to as the *ground state*. Upon measurement, the state of the entity collapses from its current state (possibly the ground state) and is projected onto one of the eigenstates.

Now consider two entities $A$ and $B$ with Hilbert spaces $\mathcal{H}_A$ and $\mathcal{H}_B$. We denote amplitudes associated with the first as $a_i$ and amplitudes associated with the second as $b_j$. The Hilbert space of the composite of these entities is given by the tensor product $\mathcal{H}_A \otimes \mathcal{H}_B$. We may define a basis $|e\rangle_i$ for $\mathcal{H}_A$ and a basis $|f\rangle_j$ for

$\mathcal{H}_B$. The most general state in $\mathcal{H}_A \otimes \mathcal{H}_B$ has the form

$$|\psi\rangle_{AB} = \Sigma_{i,j}\, cij\, |e\rangle_i \otimes |f\rangle_j \qquad (1)$$

where *cij* is the amplitude corresponding to the composite entity.

The phenomenon of *entanglement* was conceived to deal with situations of non-separability where different entities form a composite entity. The state $|\psi\rangle_{AB}$ is separable if for the amplitudes *cij* amplitudes $a_i$ and $b_j$ can be found such that $cij = a_i\, b_j$. It is inseparable, and therefore an entangled state, if this is not the case, hence if the amplitudes describing the state of the composite entity are not of a product form.[4] Entangled states are non-compositional because they may exhibit emergent properties not inherited from their constituent components.

Applications of this formalism to concepts are generalizations that have nothing to do with the quantum level of reality, thus they are sometimes referred to as “quantum-like”. In such models a context plays the role of a measurement. A set of basis states related to a context represent instances of a concept. A context can exert either a deterministic or probabilistic influence on the state of a concept.

In one generalized quantum formalism, namely the State Context Property (SCOP) theory of concepts, a concept is defined in terms of (1) its set of states $\Sigma$ (including both exemplars and ground states changed under the influence of a context), a set $\mathcal{L}$ of relevant properties, (3) a set $\mathcal{M}$ of contexts in which the concept may be relevant, (4) a function $\nu$ that gives the applicability or *weight* of a certain property for a particular state and context, and (5) a function $\mu$ that gives the probability of transition from one state to another under the influence of a particular context. We might represent the state of a lamp by a vector $|p\rangle$ of length equal to 1 in a complex Hilbert space $\mathcal{H}$. From a different context this state $|p\rangle$ could actualize as another state. For example, in the context <u>office</u> it may actualize as the state OFFICE LAMP, while in the context <u>bedroom</u> it may actualize as the state BEDSIDE LAMP. Since there is no uncertainty or choice involved, the change of state is deterministic, and this is represented by a linear operator.

If there is uncertainty involved then the change of state is probabilistic. Different possible outcomes can occur, each with a certain probability, and the effect of context is represented by a self-adjoint operator. For example, consider the reconceptualization of what to do with an old glass lampshade for which the wiring no longer works. Instances of concepts will be capitalized, instances of contexts will be underlined, and properties will be in single quotes. The concept of the lampshade is referred to as LAMPSHADE. LAMPSHADE may change probabilistically from $|p\rangle$ to one of two states: $|u\rangle$, the state in which it is viewed as useful (*e.g.*, if the individual can rewire it, or dream up a new use for it), and $|w\rangle$, the state in which it is viewed as waste. The state of LAMPSHADE prior to being conceived of as useful or waste is modeled as a superposition of these two possibilities. The vectors $|u\rangle$ and $|w\rangle$ form the basis of a complex Hilbert space. This means that any vector, including $|p\rangle$ can be written as a superposition of $|u\rangle$ and $|w\rangle$, *i.e.*,

$$|p\rangle = a_0|u\rangle + a_1|w\rangle \tag{2}$$

where $a_0$ and $a_1$ are complex numbers that give the amplitudes of $|u\rangle$ and $|w\rangle$ respectively. More concretely, for state $|p\rangle$ of LAMPSHADE, the probability it is viewed as useful equals $|a_0|^2$, the square of the absolute value of $a_0$. The probability it is viewed as waste equals $|a_1|^2$, the square of the absolute value of $a_1$. If it were decided that LAMPSHADE is useful or waste its state would change probabilistically from $|p\rangle$ to $|u\rangle$ or $|w\rangle$. The states $|u\rangle$ and $|w\rangle$ are thus eigenstates of LAMPSHADE in the context broken.

Recall that states are represented by unit vectors, and all vectors of a decomposition such as $|u\rangle$ and $|w\rangle$ have unit length, are mutually orthogonal, and generate the whole vector space; thus $|a_0|^2 + |a_1|^2 = 1$. Therefore, the change in the probability that the lamp is viewed as useful if one considers it from a different context can be modeled using a Pythagorean argument (Fig. 3).

--------------------------------

Insert Figure 3 About Here

--------------------------------

For someone else, $a_0$ and $a_1$ may have different values (as epitomized in the saying, “one person’s trash is another person’s treasure”). Referring back to the two modes of thought discussed earlier, the lampshade may also be conceived differently by the same person in different modes of thought (Fig. 1). In an analytic processing mode it may be considered useful only when properly wired to provide light. Therefore $|u\rangle = |l\rangle$ where $|l\rangle$ denotes using it to give light. An associative processing mode may enhance the capacity to view LAMPSHADE in a new context, such as turning it into art or a piece of garden décor, which changes the likelihood of it being conceived of as useful or waste. Activation of the set $\mathcal{L}$ of properties of the glass lampshade, *e.g*., 'concave', 'shallow', and 'weather resistant' denoted $f1$, $f2$, and $f3$, causes activation of other concepts for which these properties are relevant. Similarly, contexts for which some of these properties are relevant, such as garden décor, become candidate members of the set $\mathcal{M}$ of relevant contexts for LAMPSHADE.

The restructured conception of LAMPSHADE in the context garden décor, denoted $|p_g\rangle$, is given by

$$|p_g\rangle = a_3 \, Pu_g \, |p_g\rangle / \| \, Pu_g \, |p_g\rangle \| + a_4 |w_g\rangle \tag{3}$$

where $Pu_g$ is an orthogonal projection operator. We substitute in the mathematical formalism of Hilbert space for the unit vector whenever what physicists call *degeneration* is involved, meaning that several orthogonal states can amount to the same thing, here the outcome of being useful. Note that $\|Pu_g \, |p_g\rangle\|$ is the length of $Pu_g \, |p_g\rangle$. We need to divide the vector $Pu_g \, |p_g\rangle$ by $\|Pu_g \, |p_g\rangle\|$ for it to become a unit vector, and hence represent a state. Let us specify these states of usefulness. Given the context garden décor, denoted $g$, some creative states of LAMPSHADE are to use it as a birdbath or a planter. This is not possible in the Hilbert space formed by states $|u\rangle$ and $|w\rangle$ because it has only two dimensions, so we introduce the states $|b\rangle$ and $|r\rangle$ to denote BIRDBATH and PLANTER respectively. In their default context as garden décor they are denoted $|b_g\rangle$ and $|r_g\rangle$. $|l_g\rangle$ denotes the possibility that even in the context garden décor the lampshade is rewired to function as a lamp. $|l_g b_g\rangle$ and $|l_g r_g\rangle$ denote the possibility that in this context the lampshade functions as a birdfeeder and a planter, respectively. We write the projector as the sum of the partial projectors on the states. Hence we have

$$Pu_g = |l_g\rangle\langle l_g| + |l_g \, b_g\rangle\langle l_g \, b_g| + |l_g \, r_g\rangle\langle l_g \, r_g| \tag{4}$$

where $|l_g\rangle\langle l_g|$, $|l_g \, b_g\rangle\langle l_g \, b_g|$ and $|l_g \, r_g\rangle\langle l_g \, r_g|$ are the one dimensional orthogonal projection operators on the vectors $|l_g\rangle$, $|l_g \, b_g\rangle$ and $|l_g \, r_g\rangle$ respectively. The model incorporates the fact that the perceived probability that the lampshade is useful increases by considering it in different contexts, *i.e.*, $|a_3|^2 > |a_0|^2$ because the state $Pu_g\, |p_g\rangle / \| Pu_g\, |p_g\rangle \|$ incorporates eigenstates of being used not just in a lamp but as a birdbath or planter.

The set of properties of BIRDBATH includes 'leak proof' because it must hold water. This property is denoted $f_4$. The lampshade is not leak proof because it has a hole in the bottom where the wire is supposed to go through. Since 'leak proof' is not a property of the lampshade, $\nu(p, f_4) << \nu(b, f_4)$. Writing the unit vector $Pu_g|p_g\rangle / \| Pu_g|p_g\rangle \|$ as a superposition of vectors $|l_g\rangle$, $|l_g\, b_g\rangle$ and $|l_g\, r_g\rangle$ we have:

$$Pu_g|p_g\rangle / \| Pu_g|p_g\rangle \| = a_5|l_g\rangle + a_6|l_g\, b_g\rangle + a_7|l_g\, r_g\rangle \tag{5}$$

Because the lampshade is not leak proof, $|a_6|^2$ is small. However, if the lampshade were used as a PLANTER, the hole could function beautifully to drain excess water, so $\nu(p, f_4) \approx \nu(r, f_4)$. Therefore, $|a_7|^2$ is large, and $\mu(r, g, p) >> \mu(b, g, p)$. In the context garden décor, the concept LAMPSHADE has a high probability of collapsing to LAMPSHADE PLANTER.

We can model the emergence of new properties using the notion of entanglement. Although the state LAMPSHADE PLANTER was modeled by $l_g\, r_g\rangle$ as one of the sub-states of LAMPSHADE, the quantum formalism can also be used to derive this state as a combined state of LAMPSHADE and PLANTER. It has been shown experimentally that such a combined state is in general not a product state but an entangled state (Aerts, Gabora & Sozzo, 2013, Aerts & Sozzo, 2011). Thus, using the quantum formalism it is possible to model how LAMPSHADE PLANTER can actualize new properties, i.e., properties that are not properties of LAMPSHADE or PLANTER alone, such as wire-hole drainage.

This example shows that it is possible to model the creative restructuring of a concept (e.g., LAMPSHADE) when considered in a new context (e.g., garden décor). The approach provides a formal model of an incomplete idea, or what Finke, Ward, & Smith (1992), refer to as a 'pre-inventive structure' and the process of imagining a concept or idea from a seemingly incompatible point of view, referred to by Rothberg (2015) as Janusian thinking. The approach has been expanded to incorporate larger conceptual structures (Gabora & Aerts, 2009), as well as how the same concepts are conceived of differently in associative versus analytic modes of thought (Veloz, Gabora, Eyjolfson, & Aerts, 2011).

While the notion of new idea elements inducing creative shifts in mental models or problem frames as new ideas or information is introduced has been around for some time, note how different this approach is to that of, for example, Arecchi (2011), who models creativity as a "jump from one Bayesian model to another". In the quantum representation, probability is treated as arising not from a lack of information per se, but from the limitations of any context (even a 'default' context). A context that brings about an entropy-decreasing creative solution (e.g., LAMPSHADE PLANTER) may lead to another context that opens up a new problem (e.g., where should I put the LAMPSHADE PLANTER?). As in Guastello's (2002) dynamic theory of creativity, the creative thought process entails self-organization and reduces entropy. However in the HT model, prior to the realization that a wire hole could be used to drain water, it is not true that this idea "existed", nor did it "not exist"; it existed as a potentiality. Thus it is not appropriate to describe the creator's cognitive state as a collection of discrete possibilities that are searched or selected amongst. We cannot describe this state with any theory of creativity that does not incorporate potentiality, entanglement, and the emergence of new features. Finally, note how in this approach backtracking (returning to an earlier solution or idea) involves considering an ill-defined conception in different contexts as opposed to selecting amongst well-defined possibilities. A thought that (externally) gives rise to a dead end may (internally) modify the conception of the task in a way that is essential to bring it to completion. Consistent with the widespread notion that creative people are relatively unafraid of making mistakes, what appears to be a failed trial may be an indispensable step toward the final product. Trial and error may occasionally come into play, but more often creativity involves thinking something through in a nuanced and penetrating way, considering it from different angles until it comes into focus.

## Focus on Internal Transformation Rather than External Product

HT posits that creativity originates from the propensity of a worldview for self-organized restructuring resulting in a lower-energy state. Though different processes may be involved (e.g., analogy, working backward from a goal state, and so forth), they all culminate in the restructuring of worldviews and the evolution of culture; thus HT provides an integrative framework for diverse creative processes. The creative mind uses psychological entropy (arousal-provoking uncertainty) to detect and track concepts in states of potentiality, consider them from different contexts, and repeat this process on other concepts that enter states of potentiality following the original reconceptualization (such as PLANTER in the LAMPSHADE example) until psychological entropy reaches an acceptable level.

Creative outputs are viewed as byproducts of immersion in a creative task that, by making the internal process accessible, assist in the tracking and cultural transmission of creative experience. To convey an idea to others one must express it in a form they will value, and in the process of finding acceptable forms for ones' ideas the creator's worldview may achieve a better fit with societal norms. Change brought about in the worldviews of those exposed to a creative work need not mirror the creator's own experience creating it; each worldview can provide a different way of assimilating the idea and new contexts in which it may evolve further.

The internal and external effects are related; for example, self-report measures of the ability to experience original, appropriate combinations of emotions (referred to as emotional creativity) are correlated with creativity test scores (Ivcevic, Brackett, & Mayer, 2007). Moreover, imaginative pretend play and divergent thinking are correlated with emotion regulation in children (Hoffmann & Russ, 2012). However, although it is commonly assumed that the creative individual is compelled "to transform the creative idea into a creative product" (Thrash, Maruskin, Cassidy, Fruer, & Ryan, 2010, p. 470), to the intrinsically motivated creator the product may function primarily as a means of tracking the progress of the idea; its worldly value may be of secondary importance.

## Honing Perspective on Incubation and Insight

There is mixed evidence (reviewed in Sio & Ormerod, 2009) that creativity can benefit from time away from the creative task to "incubate" while engaged in other tasks. From a honing perspective, so long as psychological entropy is decreasing there is a sense of progress and incubation is unnecessary. Returning to the concept of eigenstate (which as we saw is a state that is definite with respect to some observable or context), we say that one's conception of a problem is in an eigenstate with respect to a given context when it is not subject to change. If the task is in an eigenstate with respect to every vantage-point one can think of, psychological entropy does not decrease, and there is no sense of progress. In this case incubation may be useful; the creator's worldview may shift just enough that, upon returning to the task, a context comes to mind for which the problem is not in an eigenstate.

While some attribute the unpredictability of insight to the blind, trial-and-error nature of idea generation, another explanation is that it marks a cognitive phase transition (Stephen, Boncoddo, Magnuson, & Dixon, 2009) arising due to self-organized criticality (SOC) (Gabora, 1998; Schilling, 2005).[5] SOC is a phenomenon wherein, through simple local interactions, complex systems tend to find a critical state poised at the cusp of a transition between order and chaos, from which a single small perturbation occasionally exerts a disproportionately large effect (Bak, Tang, & Weisenfeld, 1988). The signature of SOC is an inverse power law relationship between the magnitude of a critical event and the frequency of critical events of that magnitude. The hypothesis that insight is a self-organized critical event is consistent with findings that although often preceded by intensive effort, it tends to come suddenly (Gick & Lockhart, 1995) and with ease (Topolinski & Reber, 2010). SOC gives rise to structure that exhibits sparse connectivity, short average path lengths, and strong local clustering. Other indications of SOC include long-range correlations in space and time, and rapid reconfiguration in response to external inputs. There is evidence of SOC in the human brain, e.g., with respect to phase synchronization of large-scale functional networks (Kitzbichler, Smith, Christensen, & Bullmore, 2009). There is also evidence of SOC at the cognitive level; word association studies have shown that concepts are clustered and sparsely connected, with some having many associates and others few (Nelson, McKinney, Gee, & Janczura, 1998; Nelson & McEvoy, 1979; Nelson, McEvoy, & Schreiber, 2004).[6] Thus, semantic networks exhibit the sparse connectivity, short average path lengths, and strong local clustering characteristic of self-organized complexity and in particular 'small world' structure (Steyvers & Tenenbaum, 2005).

Like other SOC systems, a creative mind may function within a regime midway between order (systematic progression of thoughts), and chaos (everything reminds one of everything else). Much as most perturbations in SOC systems have little effect but the occasional perturbation has a dramatic effect, most thoughts have little effect on one's worldview, but occasionally one thought triggers another, which triggers another, and so forth in a chain reaction of conceptual change. This is consistent with findings that large-scale creative conceptual change often follows a series of small conceptual changes (Ward, Smith, & Vaid, 1997), and with evidence that power laws and catastrophe models are applicable to the diffusion of innovations (Jacobsen & Guastello, 2011). Indeed, this explanation of insight fits into a broader conceptual framework when creativity is viewed in terms of its role in fueling cultural evolution. According to the theory of punctuated equilibrium, for which there is substantial well-documented evidence, change in biological species is restricted to rare, rapid events interspersed amongst prolonged periods of stasis (Eldridge & Gould, 1972). Punctuated equilibrium is perhaps the best-known example of SOC. It has been suggested that cultural evolution exhibits punctuated equilibrium (Orsucci, 2008). Insight may play the role in cultural evolution that is played by punctuated equilibrium in biological evolution, fueling the reorganization of, not species in an ecosystem, but concepts in an 'ecology of mind.'

**The Self-made Worldview**

Knowledge is acquired through both social learning and individual learning, the later of which includes not just trial and error learning but also restructuring of existing representations. It is proposed that individuals fall along a spectrum from "self-made" to "socially-made." Individuals with a "self-made" worldview hone ideas until they form a coherent web and take a form that bears little resemblance to the form in which they were originally acquired. At the "socially-made" end of the spectrum are those whose views have changed little since they were first acquired. Since creative individuals are compelled to 'put their own spin' on what they learn, or 'make it their own' (Feinstein, 2006), over time they are more likely to end up with a self-made worldview. A self-made worldview deviates more from the status quo, so it has greater potential to produce something the world will deem creative.

A step toward a self-made worldview would be a proclivity to shift to a defocused state of attention and thereby enter an associative mode of thought. More aspects of attended stimuli participate in the process of encoding an instant of experience to memory and evoking 'ingredients' for the next instant of experience. The more they can in turn evoke, and so on. Thus, the more detail with which items have been encoded in memory the greater their potential overlap with other items, and the more retrieval routes for creatively forging relationships between current and past experiences. The individual can capitalize on fortuitous associations and thus items are considered from more perspectives. Therefore, streams of thought last longer. If something does manage to attract attention, it tends to be more thoroughly honed before settling into a particular interpretation of it. New information, including socially transmitted information, is less able to compete with the processing of previously acquired information already set in motion. The more detail with which items were encoded, the more complex the worldview, and the more routes by which one question can lead to others. Thus the longer it takes to reduce psychological entropy to an acceptable level, but the more unique the worldview eventually becomes, and the more creative the ideas may be. Once psychological entropy is reduced new ideas may not come as readily (though the self-made individual may remain creative by drawing on a reservoir of earlier ideas). Thus, there is expected to be a tradeoff between how detailed one's representations tend to be, and the age at which creativity peaks. While the ideas arising from a self-made worldview may appear unusual or risky, from the broad perspective achieved as a result of extensive honing they may be a clear move forward.

A self-made worldview may have flaws not present in a socially made worldview because it is composed of elements that have not been refined by generations of predecessors. However, from the perspective of this new way of putting together an understanding of the world, certain gaps in the human knowledge base that are waiting to be explored may be more apparent, and potentially better understood. Therefore, it is expected that, over time, individuals with a tendency toward associative thought end up with a more unique and potentially finer-grained understanding of their world. There is some evidence that creativity and intelligence are correlated up to an IQ of approximately 120, after which they diverge (Barron, 1963), although this was not supported by a recent meta-analysis (Kim, 2008). Nevertheless they appear to draw upon overlapping but different sets of executive skills (Benedek, Jauk, Sommer, Arendasy, & Neubauer, 2014; Nusbaum & Silvia, 2011). From a honing perspective it seems logical that an individual with a high IQ could master the mental skills for complex thought but possess a worldview that is not sufficiently self-made to contribute outputs that differ significantly from what has come before.

## Artistic versus Scientific / Technical Creativity

HT offers a new perspective on the distinction between artistic and scientific creativity. Artistic pursuits are often referred to as ill-defined or open-ended, and said to admit multiple "good enough" solutions rather than one "correct answer" (Kozbelt, Beghetto, & Runco, 2011, p. 33). Science appears to be more constrained than art; two scientists are more likely to converge on the same formula or gene sequence than two artists are to produce the same painting. In both cases creative individuals seek to identify gaps in their worldviews and restructure accordingly. Scientific creativity is associated with mending gaps in regions of our worldviews that are highly similar from one individual to another, that have to do with more or less universal experiences of such things as animals, gravity, machines, or lessons learned in a classroom. Artistic creativity is associated with mending gaps in regions of our worldviews that have to do with personal experience. The topology of the internal adaptive landscape associated with scientific achievements is roughly similar for everyone (though some explore it much more thoroughly) while that associated with artistic works is more unique to the individual. This suggests that although artistic projects may *seem* open-ended because the artist is expressing a region of his or her worldview to which others have less access, the artist is as constrained as the scientist and must work just as hard to "get it right". This is consistent with findings that talented artists, like scientists, have a clear sense of what constitutes a successful finished product (Feinstein, 2006). It is also consistent with evidence that the advantages associated with expertise—such as enhanced ability to remember domain-relevant patterns and generate effective problem representations—can facilitate performance even in domains such as music composition and art (Ericsson, 1999; Kozbelt, 2008). These advantages help the creator "get it right" though for artistic works others may have less knowledge of what "getting it right" entails.

**Summary of Honing Theory**

HT is founded on appreciation of the role creativity plays in fueling the evolution of cumulative, complex, adaptive culture. It is based on the proposal that what evolves through culture is minds, not external artifacts and behavior per se; though artifacts and behavior participate in their evolution and reflect the evolutionary states of the minds that generate them. Minds evolve through two mutually synergistic processes of communal exchange and self-organized restructuring, and a process is viewed as creative to the extent that it induces restructuring in one or more individuals. Minds are alerted to 'gaps'—arenas of incompletion or inconsistency in their worldviews that invite exploration—using a variable referred to as psychological entropy. Entropy is reduced by restructuring associations, of different types and at different hierarchical levels, using neural synchrony and dynamic binding. At the conceptual level this involves recognizing concepts that are in states of potentiality, and reiteratively viewing them from different contexts. Although different cognitive process are said to be creative (e.g., analogy, divergent thinking, working backward from a goal, and so forth), these processes restructure worldviews and have an impact on the evolution of culture; thus HT can provide an integrative framework for them.

Existing evolutionary theories of creativity, and the theory proposed here, are summarized in Table 2. HT posits that it is the assemblage of human worldviews that evolves over time, as did primitive life, not through Darwinian competition and survival of the fittest, but through mutual exchange and psychological entropy reducing restructuring of increasingly resilient *components*. Creative products are the manifestations of this process. As ideas are encountered by new individuals they get seen from new perspectives, enabling new avenues for context-driven actualization of potential. Creative products reflect the states of the worldviews that generate them. Our propensity for non-Darwinan cognitive change both augments our creativity and makes us vulnerable to psychiatric disorders (Barry, 2013).

------------------------------------------------

Insert Table 2 About Here

------------------------------------------------

## Application to a Creative Task

The theory of creativity presented in the previous section differs from other theories in several respects. It expands on the view that creativity is driven by a need, or gap in one's understanding, by viewing creativity as the process by which internal models of the world—worldviews—heal and reorganize themselves, thereby fueling the evolution of culture. It posits that the creative process occurs through not search or selection amongst well-defined candidates but context-driven actualization of potential. It suggests that creativity be defined and measured in terms of not external output but entropy-reducing cognitive restructuring. We now examine how HT can be applied to a particular creative achievement, using the example of the invention of the airplane.

The Wright brothers were not the first to invent an airborne machine. Out of the innumerable ways of tinkering with such a machine, how did they hit upon the successful restructuring of the problem that made their machine successful; that is, which of the infinite number of ways of relaxing constraints or elaborating a problem will lead to a solution?

The dissonance or gap in Wilbur Wright's worldview lay in the fact that he did not know how to build an airplane whereas he believed that human flight was possible. Information from seemingly unrelated domains leaked into his conception of the problem, including knowledge of bird flight, Otto Lilienthal's failed voyage, and elements of Wilbur's experience as a bicycle mechanic. The key insight in the invention of the airplane occurred to Wilbur as he idly twisted an inner tube box (Heppenheimer, 2003). Let us examine a possible scenario for how Wilbur's insight came about. Wilbur comments to Orville that when one wants to make a turn on a bicycle one banks or leans into the turn. Orville replies that birds bank into a turn by changing the angle of the ends of their wings to make their bodies roll left or right. Orville's reply has little immediate effect on Wilbur's conception of the problem because it is not obvious how to do what birds do with the solid wings of a flying machine. Wilbur's inability to solve the problem rationally leads to a spontaneous defocusing of attention. He assumes a more associative processing mode, his conception of the problem broadens, and the associative structure of his memory is probed more diffusely, though in a context-sensitive manner. As neurds get recruited his thoughts start incorporating memories of banking while riding a bicycle, and of watching birds bank into a turn.

At this moment Wilbur happens to twist and bend the inner tube box. This act becomes part of the context that interacts with his conception of the problem and causes it to shift. The set of relevant features now includes 'bends' and 'can be pulled'. A burst of alpha band activity in his right occipital cortex may have caused neural inhibition of sensory inputs, facilitating awareness of the possible relevance of these features to the problem of flight. The subsequent interaction between problem and worldview evokes memories of methods for bending solid objects. Wilber has the insight that they could twist the trailing edges of the wings in opposite directions. However, the idea is still vague because it does not include a means of accomplishing this. Note that the different ways in which the idea could materialize (e.g., by pulling cables or using a lever…) do not yet exist as coherent thoughts let alone material outcomes, so they are not in a state in which they could compete or undergo selection.

The recruiting of neurds that respond to situations involving bending and changing the direction of something long and thin by pulling on it yields an initial conception of wing-warping. The idea has some properties of bicycle steering (using hands to control banking), some properties of bird flight (banking by changing wing angle), and some properties of twisting an inner tube box (redirecting something long and thin by pulling it). The newly forged association between these properties is not literally a connection, but a distributed set of features that have never before been activated together as an ensemble. The resulting reconceptualization of the flying machine reduces the dissonance between Wilbur's confidence that flight was possible and his previous failure to make it happen.

Having arrived at an idea for how to outfit the flying machine with a steering device they had to determine if it would work in practice. Although in the short run diffuse activation is conducive to creativity, in the long run it is distracting; the relationship between one thought and the next may be tenuous, and thought lacks continuity. Therefore, now that the basic idea was in place, the brothers entered an analytic processing mode by focusing attention on the promising aspects of Wilbur's idea of applying the insight about bird wings to airplane wings (bending the tips), and ignoring irrelevant aspects (such as that bird wings can be different colors). Fewer cell assemblies are activated, and fewer features play a role in the formation of the next thought. By shifting along the spectrum from analytic to associative as needed, the fruits of one mode of thought become ingredients for the other. The Wright brothers test the idea of wing warping using kites and gliders attached to ropes controlled by sticks, and their idea is further perfected using wind tunnels and even a three-wheeled air-lifting bicycle. Each test reduces potentiality associated with the relevant concepts, and thereby decreases (at least momentarily) psychological entropy.

## EMPIRICAL SUPPORT

This section presents the results of studies that have been carried out to test predictions of HT.

**States of Potentiality**

We have seen that creative processes are generally thought to involve searching or generating a collection of discrete, well-defined possibilities, followed by selection and exploration of the most promising of them. In contrast, HT predicts that creativity involves the merging of memory items resulting in a representation that is in a state of potentiality, which undergoes honing. This section outlines studies that were conducted to test the different predictions of these different conceptions of creativity.

*Analogy Problem Solving*

We saw earlier that Gentner's (1983) structure mapping theory of analogy assumes that candidate sources are considered separately and once a source is found the analogy takes shape in isolation from the rest of the contents of mind. In contrast, according to HT, alignment and mapping work with, not individually considered, discrete, predefined items, but an amalgam of items previously encoded to the neural cell assemblies activated by the target, which may not be readily separable. The analogy making process proceeds not by mapping correspondences from candidate sources but by weeding out non-correspondences (see Fig. 4).

--------------------------------

Insert Figure 4 About Here

--------------------------------

These different theories of analogy were pitted against one another in a study in which participants were interrupted midway through solving Gick and Holyoak's (1980) classic 'The General / Radiation Problem' analogy problem and asked what they were thinking of in terms of a solution (Gabora & Saab, 2011). Data from participants who successfully completed the analogy by the time they were interrupted was discarded since we were interested in the state of a creative idea prior to completion. Naïve judges categorized a response as SM if it met the predictions of structure mapping, i.e., there was evidence that participants had searched memory and selected an appropriate source, but not finished mapping correspondences from source to target. They categorized it as HO if it met the predictions of HT, i.e., there was evidence of merging multiple possible solution sources and weeding out non-correspondences. If their answer contained extra information relative to the single complete, correct, minimally complex solution, this indicated that irrelevant non-correspondences from multiple merged sources had not yet been weeded out. (Note that for this problem there is only one complete yet minimally complex answer.) Both the frequency count and mean number of HO judgments were significantly higher than the frequency count and mean number of SM judgments.

These findings support the hypothesis that midway through creative processing an idea is in a potentiality state. Since analogy problem solving is a highly constrained task, the procedure was adapted to art-making. Naïve judges categorized artists' responses to questions about their art-making process as indicative of Theory S or Theory H using characteristics that differentiate selectionist from honing theories (Table 1). Preliminary results indicate that the findings of evidence for HT obtained with the analogy task generalize to this open-ended task (Carbert, Gabora, Schwartz, & Ranjan, 2014).

------------------------------------------------

Insert Table 1 About Here

------------------------------------------------

The results of these studies are consistent with the view that creative thinking is "divergent" not in the sense that it moves in multiple directions or generates multiple possibilities, but in the sense that it produces a raw, unfocused amalgam derived of multiple sources that undergoes restructuring through being considered from different contexts.

## Recognizable Style Within and Across Domains

Since HT posits that creative output reflects the idiosyncratic process of wrestling with issues that are in states of potentiality so as to reduce psychological entropy, the theory predicts that creative individuals may have a characteristic style or 'voice', a distinctive facet of their personality that is recognizable not just within domains but across domains. Empirical support has been obtained for this prediction (Gabora, O'Connor, & Ranjan, 2012; Ranjan, 2014). Art students were able to identify at significantly above-chance levels which famous artists created pieces of art they had not seen before. They also identified at significantly above-chance levels which of their classmates created pieces of art they had not seen before. More surprisingly, art students also identified the creators of non-painting artworks that they had not seen before. Similarly, creative writing students were able to identify at significantly above-chance levels passages of text written by famous writers that they had not encountered before, and passages of text written by their classmates that they had not encountered before. Perhaps most surprising of all, creative writing students also identified at significantly above-chance levels which of their classmates created a work of art, i.e., a creative work in a domain other than writing.

Thus individual style showed through even outside the creator's domain of expertise and even when viewers had not previously encountered works by the creator in that domain. These results are not predicted by other theories of creativity. Although it may be possible for other theories to accommodate them it is not straightforward; in particular, theories that emphasize chance or the accumulation of expertise provide no reason to expect that the works of a particular creator exhibit a recognizable style. This is, however, predicted by HT, according to which creativity is the process by which an individual's worldview is forged and expressed, and personal style reflects the uniquely honed worldview. Personal style need not be due to a conscious attempt to express style, but a side effect of participating in the human enterprise of interactively evolving cultural outputs by adapting tools and techniques to ones' tastes and perspectives in the service of expressing internal change.

The finding that creative style is recognizable across domains is incompatible with the view that creativity is domain-specific. This study differs from previous research that addresses the domain-specificity / generality question in that the focus is not on talent but on personal creative style. The results support the view that creative achievement can be characterized in terms of, not just expertise or eminence, but the ability to express what we genuinely are through whatever media we have at our disposal. This suggests that an artist's or scientist's personal style may be evident in little-c creative activities such as preparing a meal or telling a story.

## Cross-Domain Translation of Creative Ideas

Since according to HT, creative output reflects the transformative process by which an individual resolves potentiality, it predicts that it is possible for the domain-specific aspects to be stripped away such that the creative work is amenable to re-expression in another form, and recognizable in this new form. This was tested in studies that investigated the translate-ability of creative works from one domain to another (Ranjan, Gabora, & O'Connor, 2013; Ranjan, 2014). Three expert painters created four paintings, each of which was the artist's interpretation of one of four different pieces of instrumental music. Participants were able to identify which paintings were inspired by which pieces of music at statistically significant above-chance levels. This demonstration of cross-domain translation of creative ideas in conjunction with the previously mentioned evidence for the domain-generality of creativity suggest that, at their core, creative ideas are less domain-dependent than they are often assumed to be. The idea that at least some creative tasks involve the abstraction and re-expression of abstract forms seems obvious to artists, it stands in contrast with the bulk of creativity research, in which creativity is portrayed as heuristically guided search or selection amongst candidates that exist in discrete, well-defined states. The result supports the hypothesis that creative outputs reflect the idiosyncratic process by which a particular worldview reduces psychological entropy by restructures itself.

### Recognizability of Authenticity

Most theories of creativity do not address or account for the widespread belief amongst creative individuals that outputs vary in their degree of authenticity. An authentic performance is one that is natural or genuine, while an inauthentic performance feels faked, forced, or imitative. HT predicts that a creative performance will look and/or feel authentic to the extent that it reflects personally meaningful restructuring of one's worldview. Thus the theory predicts that authenticity is a real construct and that there should be agreement amongst individuals as to the degree of authenticity of creative works. This was tested in a study in which participants were asked to rate the authenticity and skill level of a series of videotaped performances by dancers and stand-up comedians (Henderson & Gabora, 2013). Authenticity ratings of the performances amongst viewers were significantly positively correlated. This finding, not predicted by other theories of creativity, suggests that authenticity is a real construct that comes through to the extent that creative performance is a revealing and accurate expression of a personal worldview.

## COMPUTATIONAL SUPPORT

Additional support for HT comes from EVOC (for EVOlution of Culture), a computational model of cultural evolution (Gabora, 1995, 2008b). EVOC consists of neural network based agents that invent new actions and imitate actions performed by neighbors. The assemblage of ideas changes over time not because some replicate at the expense of others, as in natural selection, but through inventive and social processes. Agents can learn generalizations concerning what kinds of actions are fit with respect to a particular goal, and use this acquired knowledge to modify ideas for actions before transmitting them to other agents. EVOC exhibits typical evolutionary patterns, e.g., (1) a cumulative increase in the fitness and complexity of innovations over time, and (2) an increase in diversity as the space of possibilities is explored followed by a decrease as agents converge on the fittest possibilities. EVOC has been used to model how the mean fitness and diversity of elements of culture is affected by factors such as leadership, population size and density, borders that affect transmission between populations, and the proportion and distribution of creators (who acquire new ideas primarily by inventing them) versus imitators (who acquire new ideas primarily by copying neighbors), as well as the questions pertaining to creativity reported here.

**Balancing Creative Change with Cultural Continuity**

It is commonly assumed that creativity is desirable, and that individuals, institutions, and society as a whole would be better off if everyone were more creative. However, creative individuals often claim that they feel stifled by social norms (Ludwig, 1995; Sulloway, 1996). Moreover, government policies and educational systems do not prioritize the cultivation of creativity, and in some ways discourage it (Snyder, Gregerson, & Kaufman, 2012). In short, society sends mixed messages about the desirability of creativity (Mueller, Melwani, & Goncalo, 2012).

It is possible to reconcile ideological support for creativity with societal practices that impede it by viewing creativity as a process that catalyzes the transformation of self-organizing worldviews and their evolution through culture. An evolutionary process requires a balance between change and continuity; thus it was hypothesized that, in cultural evolution, ambivalent attitudes toward creativity help maintain this balance. EVOC was used to investigate the relationship between individual creativity and the mean fitness and diversity of cultural outputs across a society (Gabora, 1995, 2008b). When agents never invented there was nothing to imitate, so there was no cultural evolution at all. If they spent even a small percentage of iterations inventing, not only was cumulative cultural evolution possible, but eventually all agents converged on optimal outputs. At the other extreme, when all agents invented every iteration instead of imitating, the mean fitness of outputs was also sub-optimal because fit ideas were not diffusing through the artificial society. The society as a whole performed optimally when the ratio of creating to imitating was at an intermediate value of approximately 2:1 (Fig. 5). The finding that the mean fitness of innovations across the artificial society increased when at least a fraction of iterations were spent imitating is consistent with the finding obtained in real groups that enabling members to compare and imitate solutions facilitates not just scrounging but also innovation, through the propagation of viable solutions for further cumulative exploration (Wisdom & Goldstone, 2011).

--------------------------------

Insert Figure 5 About Here

--------------------------------

Interestingly, for the agent with the fittest actions, action fitness was directly correlated with the creation-to-imitation ratio. This suggested the society as a whole might perform even better if there were a division of labor between creators and conformers. This was investigated using a version of EVOC that incorporates individual differences in creativity (Gabora & Firouzi, 2012). There were two types of agents: imitators that only obtained new actions by imitating neighbors, and creators that obtained new actions either by inventing or imitating. Whereas a given agent was either a creator or an imitator throughout the run, the proportion of creators inventing or imitating in any given iteration fluctuated stochastically. We found that for optimal mean fitness of agents' actions across the society there was a tradeoff between $C$, the proportion of creators to imitators, and $p$, how creative the creators were (Fig. 6). Creative agents, while essential, were interfering with the propagation of proven effective ideas, and prone to reinventing the wheel. These results provided further support for the hypothesis that society functions best when creativity is tempered with continuity.

--------------------------------

Insert Figure 6 About Here

--------------------------------

These findings suggested that in real human societies individuals adjust how creative they are in accordance with their perceived creative success. This was modeled by enabling EVOC agents to self-regulate their level of creativity in response to the fitness of their invented actions (Gabora & Tseng, 2014). Over a run they became more pronouncedly either inventors or imitators, and the mean fitness of actions was significantly higher in runs in which agents self-regulated than in runs in which they did not.

These findings, which arose from the cultural evolutionary framework of HT, support the hypothesis that society's ambivalent attitude toward creativity holds adaptive value. Although results obtained with a simple computer model may have little bearing on complex human societies, the finding that a division of labor between creators and conformers benefited the society as a whole is consistent with empirical findings of such a division in human populations (Florida, 2002). The finding that EVOC societies benefit when agents increase or decrease how creative they are in accordance with perceived creative success is consistent with the existence of pronounced individual differences in creativity (Wolfradt & Pretz, 2001), and with empirical findings of self-organization and emergent leadership in groups engaged in creative problem solving (Guastello, 1998).

The finding that the mean fitness of ideas in EVOC was highest when (1) agents imitated as well as invented, and (2) there was a mixture of creative and uncreative agents, as well as the trade-off between the proportion of creators to conformers, and how creative the creators were), support a hypothesis derived from HT's cultural evolution framework for the rarity of extremely creative individuals. Cultural evolution, like any evolutionary process, must balance change and continuity. While creative individuals generate the novelty that fuels cultural evolution, absorption in their creative process impedes diffusion of proven solutions, effectively rupturing the fabric of society. Thus a society in which creative (novelty injecting) individuals are interposed with conforming (continuity maintaining) individuals ensures both that new ideas come about and if effective they are not easily lost by society as a whole.

**Biological Evolution of the Capacity for Cumulative, Open-ended, Creative Culture**

EVOC was also used to investigate two hypotheses concerning what kind of cognitive architecture can support the uniquely human cultural evolution of cumulative, adaptive, open-ended novelty.

*Effect of Chaining*

It has been hypothesized that cultural evolution was made possible by onset of the capacity for one thought to trigger another recursively leading to the progressive embellishment and modification of thoughts and actions (Donald, 1991; see Corballis, 2011 for a related proposal). This was tested by comparing runs in which agents were limited to single-step actions to runs in which they could chain ideas together to generate multi-step actions (Gabora, Chia, & Firouzi, 2013). Chaining increased the mean fitness (Fig. 7a) and diversity (Fig. 7b) of actions in the artificial society. While chaining and no-chaining runs both converged on optimal actions, without chaining this set was static, but with chaining it was in constant flux as ever-fitter actions were found. With chaining there was no ceiling on mean fitness of actions, and chaining also enhanced the effectiveness of the ability to learn trends.

--------------------------------

Insert Figure 7 About Here

--------------------------------

These findings support the hypothesis that the ability to chain representations together can transform a culturally static society into one characterized by open-ended novelty.

*Effect of Contextual Focus*

The hypothesis that restructuring is aided by *contextual focus* (CF) was also tested with EVOC (Gabora, Chia, & Firouzi, 2013). When the fitness of an agent's outputs was low it temporarily shifted to a more associative mode by increasing $\alpha$: the degree to which a newly invented idea deviates from the idea upon which it was based. Two forms of CF were implemented. The analytic mode was implemented the same way in both, but the associative mode was implemented differently. In CF1, new ideas were generated by randomly varying known ideas. In CF2 the generation of new ideas was biased by learned generalizations. Both mean fitness (Fig. 6c) and diversity (Fig. 6d) of actions increased with CF, as hypothesized, and CF was particularly effective when the fitness function changed, which supports its hypothesized utility in adapting to new or changing environments. CF2 was more effective than CF1, which supports the hypothesis that associative thought works not by generically expanding the sphere of associates but through more pronounced consideration of context.

These findings further establish the value of a cultural evolutionary framework for creativity, and in particular one that stresses the role of creativity in fueling cultural change through structuring. Restructuring is a complex form of chaining that entails reiteratively re-examining a task from different perspectives that present themselves owing to the mind's self-organizing capacity, and CF synchronizes the extent of conceptual cross-fertilization with its potential payoff. Note that although chaining made the variety of novel outputs open-ended, and this became even more pronounced with CF, these novel outputs were nonetheless predictable. Chaining and CF did not open up new cultural niches in the sense that, for example, the invention of cars created niches for the invention of things like seatbelts and stoplights. EVOC in its current form could not solve *insight problems*, which require restructuring the solution space (Boden, 1990); nonetheless it is sufficient to illustrate the effectiveness of chaining and CF.

## GENERAL DISCUSSION

Restructuring is generally thought to occur through divergent thought, which is said to involve generically widening the sphere of associates to generate more discrete possibilities. This cannot explain why thinking creatively about tires in the context 'playground equipment' may generate TIRE SWING while thinking creatively about tires in the context 'harsh road conditions' will not. It was proposed that creative thought is "divergent" in a different sense: in the sense of producing ideas that involve uncertainty, which attract attention by generating arousal. Arousal is reduced by considering them from different contexts. Contextually relevant items affect similar regions of memory because, though perhaps seemingly different, they share properties or relations, and are thus more likely than chance to restructure the problem and generate a solution that is relevant to the task at hand, perhaps in a previously unnoticed but useful way. The occasional extensive rapid restructuring may involve SOC, and be experienced as insight.

Leakage of information from outside the problem domain occurs in a manner that is unpredictable because it involves unearthing associations that exist due to overlap of items in memory that may not have been previously noticed. Since memory is distributed, coarse-coded, and content-addressable, items encoded previously to neurds are superficially different from the present situation yet share aspects of its deep structure. Therefore, recruitment of neurds may foster associations that are seemingly irrelevant yet potentially vital to a creative idea. By responding to abstract or atypical features of the situation, neurds effectively draw new concepts into the conceptualization of the problem. This can result in emergent properties, i.e. gain or loss of features, or the formation of new concepts.

HT incorporates the noncompositional manner in which concepts interact, and the possibility that such interactions result in emergent properties, making use of a quantum-like formalism in which variables are natively context specific, which was expressly formulated to describe these phenomena using superposition, entanglement, and interference. Moreover, the quantum approach is not tacked on *ad hoc* but fits naturally within a HT framework, which places interaction amongst concepts at the heart of the creative process, i.e., it proposes that a worldview achieves a lower-energy state by exploiting problems or issues with the potential to yield new concept combinations through creative restructuring.

We said that the creative process reflects the tendency of a worldview to mend itself as does a body when injured, and thereby achieve a lower entropy state. A factor that distinguishes this theory of creativity from others is that it focuses on restructuring not just as it pertains to the task, but to the worldview as a whole. When faced with a creatively demanding task there is an interaction between the conception of the task and the worldview. The conception of the task changes through interaction with the worldview, and the worldview changes through interaction with the task. The creative process reduces dissonance between the problem and other elements of the worldview.

Since HT involves the mind as a whole is in a better position than theories that focus exclusively on the problem domain to explain why creativity can be intrinsically rewarding (Gruber, 1995; Kounios & Beeman, 2014; Martindale, 1984), correlated with positive affect (Hennessey & Amabile, 2010), and often therapeutic (Barron, 1963; Forgeard, 2013), and can enhance ones' sense of self (Garailordobil & Berrueco, 2007; MacKinnon, 1962). Although the intrinsically rewarding nature of creativity is self-evident from, for example, the fact that people often accept low paying but creative jobs, theories of creativity that focus on output, or assume creative tasks to be carried out in a domain-specific manner in isolation from the rest of the creator's mind, are hard-pressed to explain this aspect of creativity.

Creative restructuring may reduce dissonance, unify previously disparate findings, or facilitate the identification and expression of repressed emotion, and these changes may in turn have other psychological implications. The transformative effect may operate across domains, i.e., one may obtain an understanding of ecological webs by painting murals of them.

The internal effects of creativity may also stem from its capacity to enhance feelings of connection to, and appreciation by, others. In the verification phase the creator considers the product from the context of a typical individual who will encounter it. For an inventor this might involve developing a working prototype. For an artist it might involve arranging to show paintings at a gallery. By finding a form for the idea that is palatable and comprehensible to others, one's own worldview merges with and expands the worldviews of others. To the extent that a creative product responds to universal features of worldviews it may have a healing effect on others. Creative products are felt to be a highly personal form of self-disclosure, and self-disclosure has therapeutic value (Pennebaker, 1997), and even beneficial effects on the immune system (Pennebaker, Kiecolt-Glaser, & Glaser, 1998). Since creative people often feel disconnected from others because they defy the crowd (Sternberg & Lubart, 1995; Sulloway, 1996), the benefits of creative self-disclosure may be mediated by an enhanced sense of belonging. In short, the transformative effects of creative honing on self-concept and wellbeing extend well beyond the domain of creative practice.

## Implications for Distinction between Creativity of Humans and Other Species

Although many other species engage in acts that could be called creative (Reader & Laland, 2003), their achievements are comparatively brittle and stereotyped (see Penn, Holyoak, & Povinelli, 2008 for overview). The issue is controversial (Kaufman & Kaufman, 2015), but most scientists probably agree that no other species comes close to the multifaceted, cumulative creativity of humans. While creativity is observed in other species, we are unique in the extent to which we build on each others' creative outputs, adapting them to our needs and tastes. If creativity is beneficial and adaptive, what has stopped other species from becoming remotely as creative as humans?

Penn et al. (2008) suggest that while many species use neural synchrony to code contextual associations only humans use it for the kind of dynamic binding amongst roles, fillers, and structured relations necessary to support higher-order cognition. HT is consistent with this, but posits that the uniqueness of human top-down processing originates in autopoietic structure and the use of psychological entropy to drive emotions and intuitions that guide creative honing. Much as the ingredients for life do not guarantee life, the capacity for dynamic-binding among roles, fillers, and structured relations does not guarantee that representations coalesce into a web of understandings that is (1) sufficiently integrated to prioritize, adapt plans to new circumstances, and creatively express oneself, yet (2) sufficiently compartmentalized that only those associations that are meaningful and relevant to the current context come to mind. Recall from the discussion of Kauffman's simulations of autocatalytic closure earlier in this paper that each polymer was composed of up to a maximum of $M$ monomers, and assigned a low random probability $P$ of catalyzing each reaction. $P$ had to fall within a narrow range of values in order for autocatalytic closure to occur, and the lower the value of $P$, the greater $M$ had to be, and *vice versa*. A similar trade-off is expected between the $M_C$ (the cognitive analog of Kauffman's $M$), the maximum number of dimensions along which representations are encoded, and $P_C$ (the cognitive analog of Kauffman's $P$), the probability that one item in memory evokes another (Gabora, 1998). So long as $\{M_C, P_C\}$ are sufficiently high, by engaging in self-sustained streams of thought that alter one's web of understandings, conceptual closure is established. At this point the associative network is integrated in that any one representation is reachable by way of some possible train of thought starting from any other representation. Once the density of associations is sufficiently high that ideas can be adapted to one's needs, tastes, and perspectives, one can contribute to cultural evolution. This hypothesis is in principle testable in a cognitive architecture in which it is possible to tune $M_C$ and $P_C$, and thereby (1) forge associations through a mechanism such as neural synchronization during associative thought, and (2) approximate the higher-order, systematic, relational capabilities of a physical symbol system during analytic thought. Bipedalism and modifications to the vocal tract enabling speech may have provided sufficient means of *implementing* complex ideas to make it worthwhile to *have* them. If so, humans are the only animal for which there has been sufficient evolutionary pressure to tinker with $\{M_C, P_C\}$ until they reliably fall within the regime that allows conceptual closure to be established.

## Limitations and Future Directions

The centrality of creativity in human cognition, and the fact that it has lagged behind other areas of cognition in terms of theoretical models, make it an important and timely topic. The proposed theory of creativity is speculative, and some of the evidence comes from a highly artificial computer model. The model is useful for manipulating variables that cannot be manipulated in the real world and exploring whether particular aspects of the theory are feasible. Although the results do not necessarily generalize beyond the model they are consistent with empirical data. In addition, HT integrates a body of knowledge from related disciplines, and provides answers to questions left unanswered by other theories of creativity.

HT predicts that creative individuals should experience longer and/or more frequent bouts of honing or abstract, associative thought, and respond to the presence of vague, ambiguous, or unusual stimuli with honing. This could be tested through self-report measures (what are you thinking about now?) or measurements of latency to stop attending ambiguous stimuli and exhibit facial expressions associated with honing.

An intriguing speculation suggested by this theory is that the psychological trait of being comfortable with ambiguity, which is strongly associated with creativity, is a high-level consequence of allowing one's concepts to remain in superposition states until they "anneal" into a low-entropy state instead of forcing them into a state that contains incompatibilities (such that the worldview could be said to be 'fractured'). A possible side effect of the creative individual's acceptance of ambiguity is less inclination to come across to others as clearly advocating particular beliefs or positions, or exemplifying particular qualities or characteristics, which may open the doors to others projecting their own issues onto the creative individual in a manner akin to a Rorschach test. This hypothesis could be tested by seeing if there is a correlation between creativity and the tendency to be seen by others as exhibiting their own faults or struggling with their own issues.

HT has implications for the assessment of creativity. The predictive validity of divergent thinking tests (while greater than that of IQ tests) is only moderate (Kim, 2008), perhaps because they involve the generation of multiple off-the-cuff solutions that constitute full-fledged solutions from the start. Creative individuals may do poorly on such tasks because they do not seem personally meaningful. In contrast, real world creative tasks often involve forming combinations that are initially incomplete, and that require honing and restructuring. This strengthens the argument for tests that involve real world creative tasks, such as Amabile's (1982) Consensual Assessment Technique.

Another direction for future research would be to investigate whether the continual bombardment of information in modern society interferes with the capacity to actualize creative potential by honing the information one already has. In-depth creative processing may be particularly needed during periods of rapid cultural change. Our challenge may be, not to generate creative ideas or find creative solutions to problems, but to come up with ideas and solutions that do not generate other problems, that take the broader context—including its social, cultural, and ecological components—into account. A theory of creativity that incorporates the worldview in which creative thinking takes place may thus constitute a step toward not just a richer understanding of how ideas come about but a more sustainable world.

## Summary and Conclusions

This paper has proposed a theory of creativity, HT, that has its roots in the question of how ideas evolve over time and what kind of structure participates in the evolution of novel forms., According to HT:

(1) The creative process fuels the evolution of culture amongst communal exchanging minds that are self-organizing, self-maintaining, and self-reproducing.

(2) Minds, like other self-organizing systems, modify their contents and adapt to their environments to minimize entropy. Creativity begins with detection of high psychological entropy material, which provokes uncertainty and is arousal-inducing.

(3) Creativity involves recursively considering this material from new contexts (i.e., honing it) until it is sufficiently restructured that arousal dissipates.

(4) A creative work may similarly induce restructuring in others, and thereby contribute to the cultural evolution of more nuanced worldviews.

HT is a non-Darwinian evolutionary theory, i.e., although it emphasizes the role of creativity in cultural evolution, it does not rely on the claim that culture evolves through a selectionist process. Since lines of cultural descent connecting creative outputs may exhibit little continuity, it is proposed that cultural evolution occurs not at the level of the outputs but at the level of self-organizing minds; outputs reflect their evolutionary state. Thus what evolves through culture is the individuals' spontaneously self-organizing internal model of the world, or worldview, and creativity is the psychological entropy-minimizing process by which self-organization occurs. The theory is consistent with evidence that creative restructuring involves neural synchrony and dynamic binding, and may be facilitated by temporarily shifting to a more associative mode of thought. HT emphasizes the internal transformative effect of creative restructuring over the external outcome. Defining and measuring creativity in terms of, not how external outputs stack up against the social yardsticks of the day, but the transformative effect of creative engagement and honing, is consistent with common attitudes toward creativity, e.g., while an original masterpiece—which provides a new roadmap for human understanding—is thought to be creative, a reproduction or imitation is not.

It was proposed that the mind's self-organizing capacity originates in a memory that is distributed, content addressable, and sufficiently densely packed that for any one item there is a reasonable probability it is similar enough to some other item to evoke a reminding of it, thereby enabling refinement of ideas and actions in a stream of thought. The more detail with which items are encoded in memory, the greater their potential overlap, and the more routes by which one can make sense of the present in terms of the past, adapt old ideas to new situations, or creatively express oneself. The capacity to string items together in a stream of thought, and to shift between associative and analytic modes of thought were crudely modeled and found to be essential for the open-ended cultural evolution of creative novelty.

Because of the architecture of associative memory, creativity involves not searching and selecting amongst well-formed idea candidates but amalgamating and honing information from multiple sources.

While creativity does involve ordinary processes such as planning and remembering, it also involves cognitive states unique to creativity which can be formally described as superposition states, and processes by which the task representation changes through interaction with an internally or externally generated context, referred to here as honing, until psychological entropy is acceptably low. The unborn idea can be said to be in a 'state of potentiality': it could actualize different ways depending on the contextual cues taken into account as it takes shape. The formalisms used by HT to describe concept combination, which lies at the heart of the creative process, are compatible with the difficult to explain findings of noncompositionality and emergent properties**.** People respond to concept combinations in ways that exhibit the predicted contextual effects as well as superposition, entanglement, and interference (Aerts, 2009; Aerts, Broekaert, Gabora, & Veloz, 2012; Aerts, Gabora, & Sozzo, 2013; Aerts & Sozzo, 2011; Bruza et al., 2008, 2011; Busemeyer & Bruza, 2012; Nelson & McEvoy, 2007). This view of creativity is consistent with research showing that midway through the creative process an idea may exist in a 'half-baked' state, in which irrelevant as well as relevant elements originating from diverse memory sources may be merged, and which midway through the creative process may not be completely comprehensible even to the creator. Other predictions of HT were supported in studies of the recognizable style both within and across domains, the cross-domain translation of creative ideas, the recognizability of authenticity, and experiments using an agent based model in which elements of culture evolve through the invention of new ideas and the imitation of neighbours.

It is hoped that this complex systems framework will stimulate further investigation of creativity, a topic that has long proven a challenge but that is central to our humanness.

## ACKNOWLEDGEMENT

This research was supported in part by a grant (62R06523) from the Natural Sciences and Engineering Research Council of Canada.

## ENDNOTES

1. The ten-year rule, though relatively robust is complicated by factors such as that the duration of study is inversely related to creative performance (Simonton, 2014).

2. Dissonance may be reduced not by revising poor ideas but weaving stories about those who point out flaws in one's ideas, yet this too is a creative, self-maintaining process.
3. It is more complex but more accurate to define a Hilbert space as a real or complex inner product space that is a complete metric space with respect to the distance function induced by the inner product. The inner product allows one to define vector length, the angle between vectors, and orthogonality between vectors (zero inner product).
4. Aerts (2009) argues that quantum field theory, which uses Fock space, is superior to the tensor product for modeling concept combination. Fock space is the direct sum of tensor products of Hilbert spaces, so it is also a Hilbert space.
5. Schilling pays lip service to BVSR but this is not integral to her theory.
6. This has culminated in a map of the conceptual distances, or degree of relatedness, amongst concepts, a feat analogous to mapping the genome for the conceptual structure of the mind.

**Table 1:** Characteristics used by naïve judges to categorize artists' responses to questions about their art-making process as indicative of Theory S or Theory H.

| **Characteristic** | **Theory S** | **Theory H** |
|---|---|---|
| If multiple ideas are given, they are | Distinct (e.g., complete ideas separated by 'or') | Jumbled together (e.g., idea fragments spliced together) |
| Ideas are | Well-defined; need to be tweaked and selected amongst | Ill-defined; need to be made concrete; later elements emerge from earlier ones |
| Common core to ideas? | Never | Yes or sometimes |
| Emergent properties? | No | Yes |
| Emergent self-understanding? | No | Yes |
| Emphasis | External product | Internal transformation |

**Table 2.** Evolutionary theories of creativity.

| **Theory of Creativity** | **Evolutionary Mode** | **Generation of Variation** | **Examples** |
|---|---|---|---|
| Early Blind Variation and Selective Retention | Darwinian | Blind | (Campbell, 1960; Simonton, 1999; Jung, Mead, Carrasco, & Flores, 2013) |
| Later Blind Variation and Selective Retention | Unclear | Partially sighted | (Simonton, 2007, 2011) |
| Evolutionary-Predictive Framework | Allegedly Darwinian | Partially sighted | (Dietrich & Haider, 2015) |
| Honing Theory | Lamarckian | "Insighted" | (Gabora, 2005) |

## FIGURE CAPTIONS

Fig. 1 (a) Evolution through natural selection. (b) Evolution with transmission of acquired traits. The significance of the different rows in (b) is identical to that in (a) but the explanations have been removed in order to make the figure less cluttered. Three dots are used to represent that many more changes may be acquired than can be indicated in such a figure. (From Gabora & Kauffman, in press).

Fig. 2 A schematic illustration of how creativity hinges on the potentiality of concepts. The *situation* of having to decide what to do with an old glass lampshade activates the concept LAMPSHADE. (a) *Analytic thought* (upper left) activates highly typical or salient properties only, represented by small filled circles, and context plays relatively little role. (b) *Divergent thought* as it is generally construed (upper right) is said to activate not just highly typical properties but also less typical or peripheral properties, represented by triangles and diamonds. *Context-driven divergent thought, sometimes called 'associative thought'* (below) activates both highly typical properties and context-relevant peripheral properties. (c) One *perspective* on the old lampshade situation (lower left) may be to view it as waste, and dispose of it. This activates one set of peripheral properties. (d) A different perspective (lower right) may be lead to seeking a creative use for it. This activates different set of peripheral properties.

Fig. 3 Graphical depiction of a vector $|p\rangle$ representing the state of LAMP. In the default context, the state of LAMP is likely to collapse to the default projection vector $|w\rangle$ which represents that it is waste. This can be seen by the fact that subspace $a_0$ is smaller than subspace $a_1$. Thinking more creatively about the LAMP one might consider many possible uses for it in the garden, for example as a birdbath or planter. Thus in the context of garden décor (shown in gray), the state of LAMP is likely to collapse to the orthogonal projection vector $|u\rangle$ which represents that it is useful, as shown by the fact that $b_0$ is larger than $b_1$. Also shown is the projection vector after renormalization (the vertical arrow).

Fig. 4 Highly simplified illustration of the differences between analogy solving by structure mapping versus honing. According to structure mapping, midway through the analogy making process a correct source has been found but not all the correspondences have been mapped. According to honing, midway through the analogy making process potentially relevant sources have been amalgamated and not all irrelevant correspondences have been weeded out. Note that once the analogy is complete it is not possible to distinguish between the two theories.

Fig. 5 Mean (a) fitness of outputs across all agents in the artificial society over runs with different creation-to imitation ratios (that is, *p(create)* values). Within a run *p(create)* was always the same for all agents. When agents only imitated (*i.e*., *p(create)* = 0), new ideas never arose and fitness of outputs never improved. However, when agents only invented (*i.e*., *p(create)* = 1), fitness was also suboptimal. The fitness of outputs across society as a whole was highest when the ratio of creating to imitating was between 0.5 and 0.75, *i.e*., approximately 2:1. (From Gabora, 1995.)

Fig. 6 The effect of varying the percentage of creators, *C*, and how creative they are, *p*, on the mean fitness of ideas in EVOC (from Gabora, & Firouzi, 2012). 3D graph (left) and contour plot (right) for the average mean fitness for different values of *C* and *p* using Present Innovation Value (PIV) discounting to ensure that the present value of any given benefit with respect to idea fitness diminishes as a function of elapsed time before that benefit is realized. The z-axis is reversed to obtain an unobstructed view of surface; therefore, lower values indicate higher mean fitness. The red line in the contour plot shows the position of a clear ridge in fitness landscape indicating optimal values of *C* and *p* that are sub-maximal for most {*C*, *p*} settings, *i.e.,* a tradeoff between how many creators there are and how creative they are. The results suggest that excess creativity at the individual level can be detrimental at the society level because creators invest in unproven ideas at the expense of propagating proven ideas. The same pattern of results was obtained analyzing just one point in time (not shown), and using a different discounting method (not shown).

Fig. 7 Above: fitness (a) and diversity (b) of outputs with and without the capacity to chain ideas together and thereby generate multi-step outputs. Below: fitness (c) and diversity (d) of outputs with and without contextual focus: the capacity to shift between a divergent processing mode when idea fitness is low and an analytic processing mode when idea fitness is high. The analytic mode works identically in CF1 and CF2. In CF1 the divergent mode operates by exploring the space of possibilities at random. In CF2 it operates by exploring the space of possibilities more conservatively, capitalizing on learned associations.

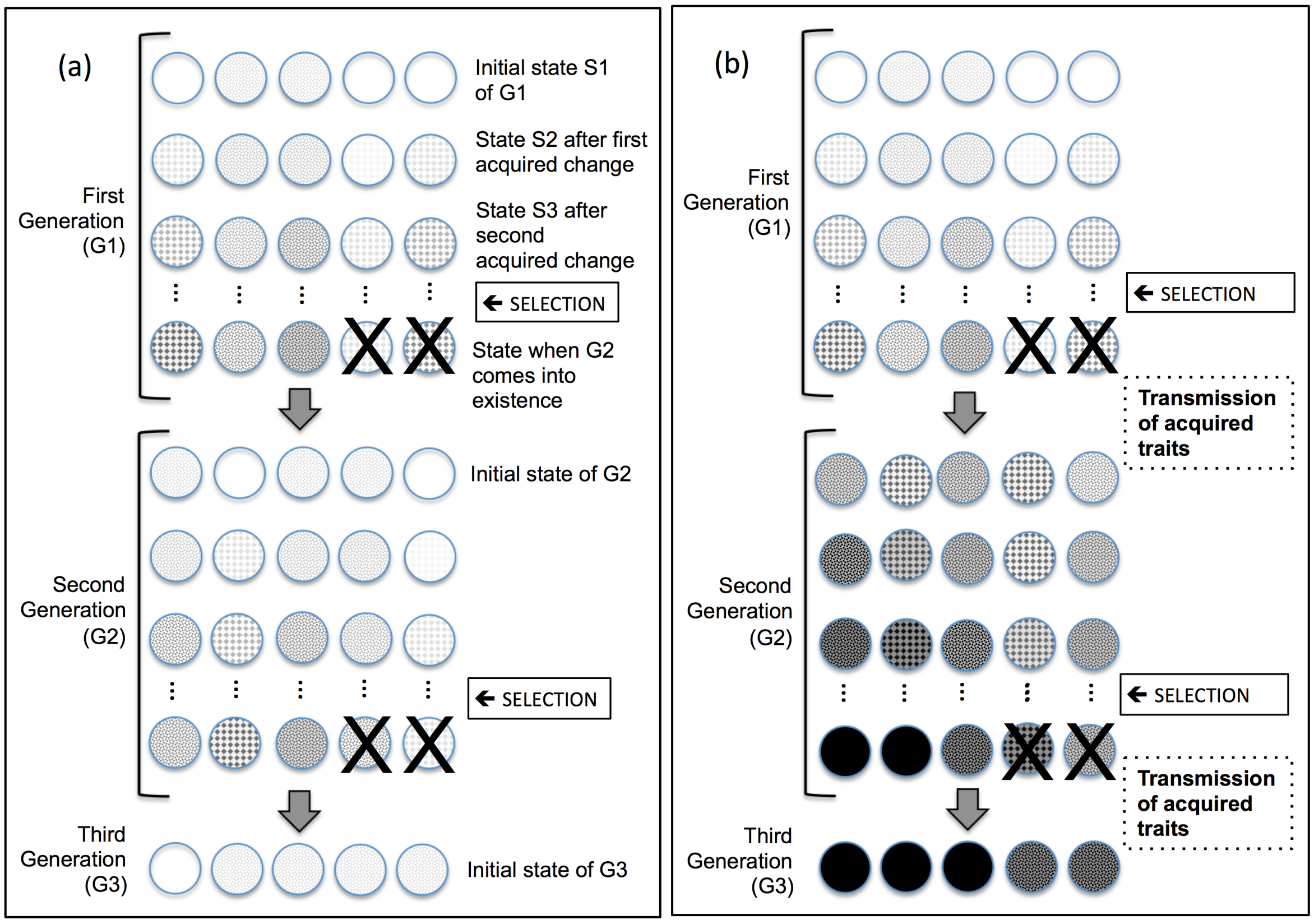

Fig. 1

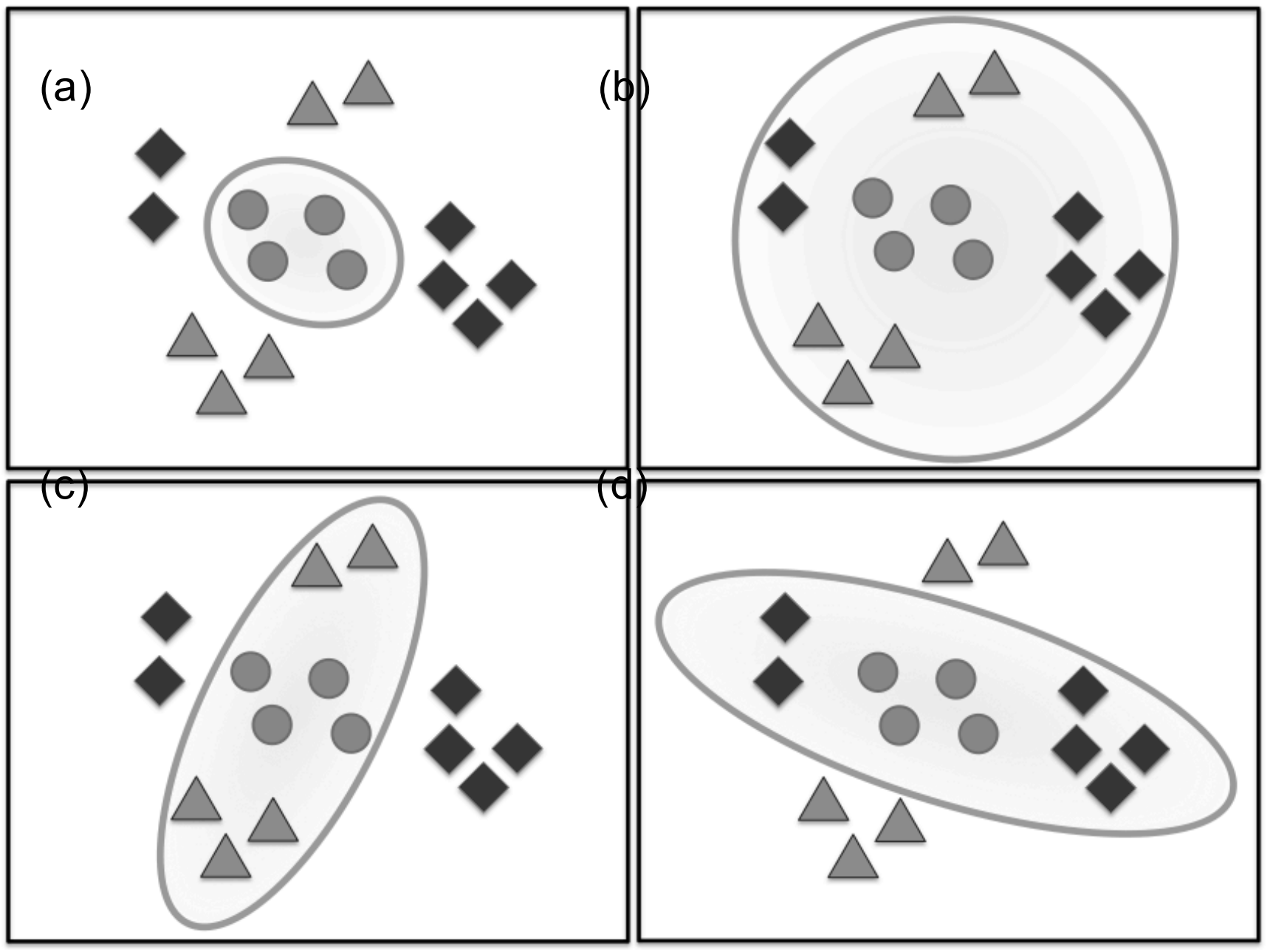

**Fig. 2**.

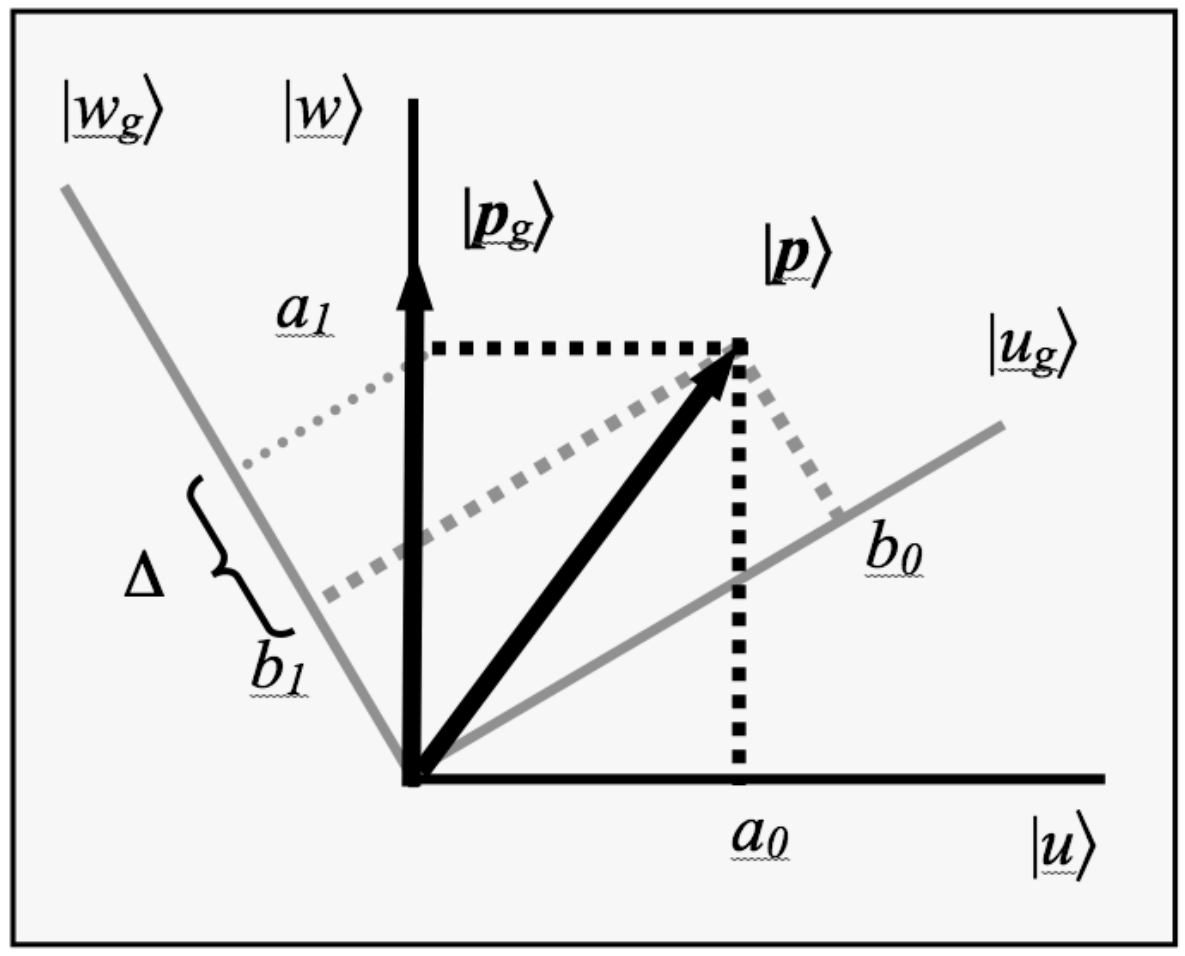

**Fig.** 3.

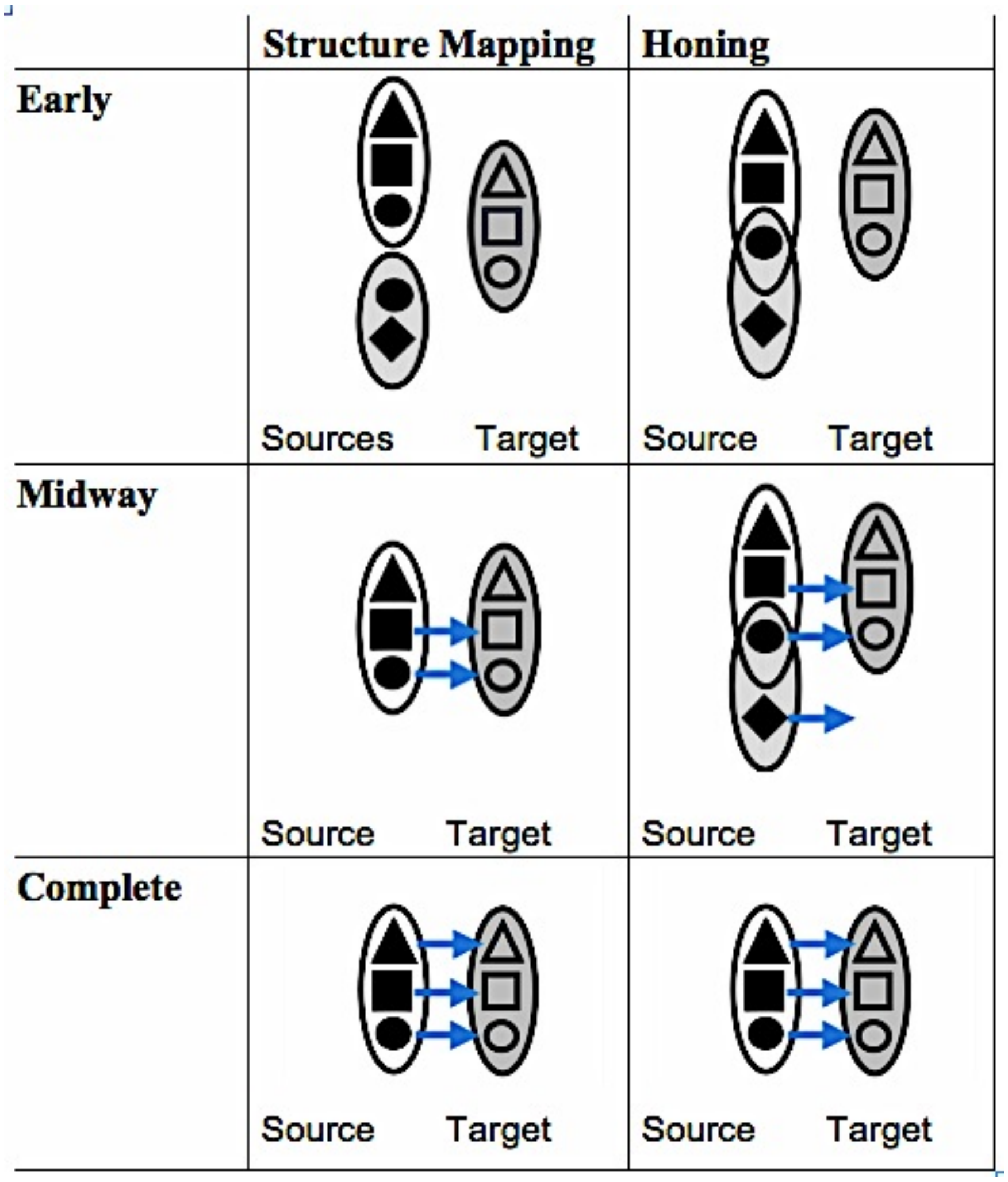

**Fig. 4.**

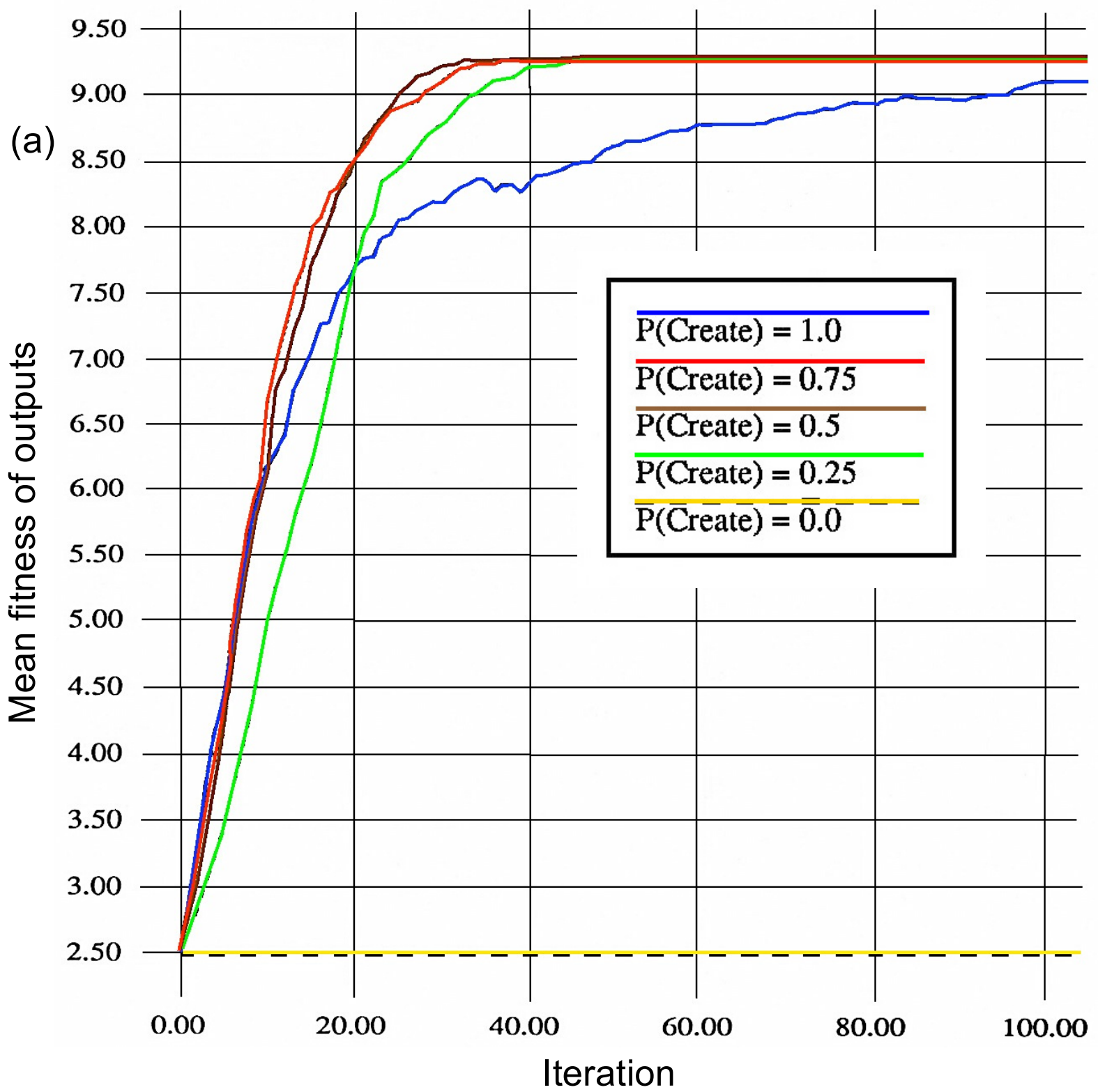

**Fig. 5.**

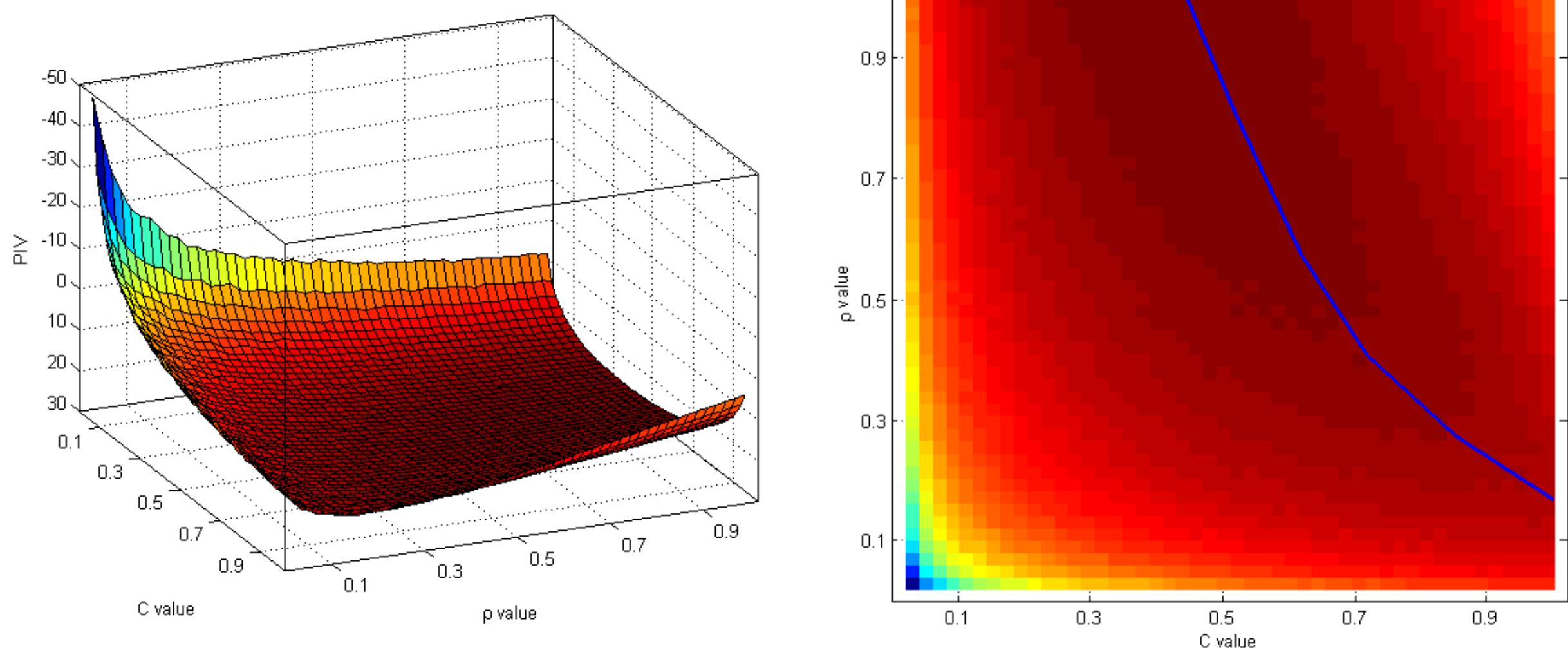

**Fig. 6.**

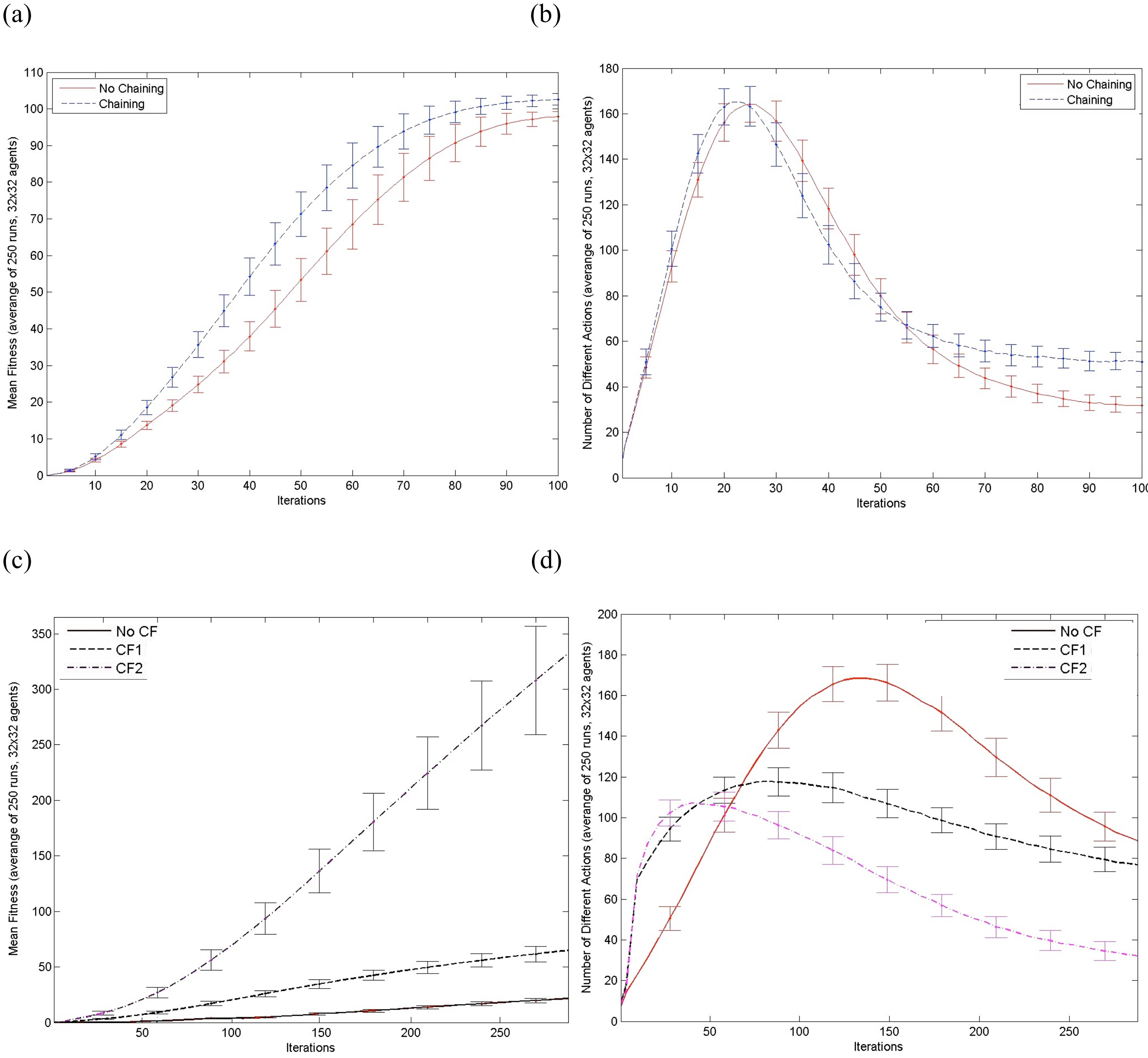

**Fig. 7.**